\documentclass[letterpaper,10pt]{article}
\usepackage{osameet3}
\usepackage[font=small]{caption}
\usepackage{graphicx}
\usepackage{subcaption}
\usepackage{amsmath,amssymb}
\usepackage[pdftex,colorlinks=true,bookmarks=false,citecolor=blue,urlcolor=blue]{hyperref} 
\newcommand{\blk}{\color{black}}
\definecolor{ngreen}{rgb}{0.2,0.6,0.2}

\definecolor{npurple}{rgb}{0.8,0.2,0.8}

\begin{document}
\title{Analytic Expressions for Orbital Angular Momentum Modal Crosstalk in a Slightly Elliptical Fiber}
\author{Ramesh Bhandari}
\address{Laboratory for Physical Sciences, 8050 Greenmead Drive, College Park,  Maryland 20740, USA}
\email{rbhandari@lps.umd.edu}
\begin{abstract}
 Assuming weakly guiding approximation, we examine orbital angular momentum (OAM) mode mixing on account of ellipticity in a fiber and derive a complete set of analytic expressions for spatial crosstalk, using scalar perturbation theory that incorporates fully the existing degeneracy between a spatial OAM mode and its degenerate partner characterized by a topological charge of opposite sign. These expressions, consequently, include an explicit formula for calculating the $2\pi$ walk-off length over which an input OAM mode converts into its degenerate partner, and back into itself. We further \blk explore \blk the applicability of  the derived expressions in  the presence of spin-orbit interaction. The expressions constitute a useful mathematical tool in the analysis and design of fibers for spatially-multiplexed mode transmissions. Their utility is demonstrated with application to  a few mode and \blk a  multimode step-index fiber.
\end{abstract}
\ocis{Orbital angular momentum, fiber ellipticity, fiber optics, waveguides, mode mixing, crosstalk, perturbation theory, analytic expressions, spin orbit interaction}


\section{Introduction}
Currently there exists a vast interest in exploring the possibility of increasing traffic flow multifold  on a fiber simply by exploiting the orthogonality of the orbital angular momentum (OAM) modes \cite{willner}. However, any such effort  must take into account fiber imperfections, like ellipticity, which can cause mode-mixing (crosstalk) and reduce the quality of the signal. 
An analytic treatment of the effect of ellipticity on OAM modes was earlier provided \cite{alex}, but it was limited to a  low topological charge $l$, primarily $l=1$ or $2$ ($l$ is the OAM carried by a single photon in units of the reduced Planck's constant ($h/(2\pi)$); besides, the methodology as presented could not incorporate mode mixing with the neighboring  OAM modes differing in topological charges by a unit or more.  More recently, several authors have carried out  numerical studies of the impact of ellipticity on OAM modes in a variety of fibers: ring fiber \cite{yue, rochelle, guerra}, multi-ring fiber with trench \cite{wang1}, and a graded-index fiber \cite{wang2} but the studies are based on computationally-intensive finite-element approximation methods, where it is not always easy to ascertain the inherent behavior of the impact of a perturbation like ellipticity in any detailed manner due to the associated high demands of accuracy.  
\\\\
In this work, we assume the weakly-guiding approximation, which permits us to write the  vector modes as products of the spatial OAM modes and the associated polarization \cite{bhandari, ramachandran}.  Further, confining our attention here to the effect  of ellipticity on the spatial mode component only, we employ the scalar wave equation and derive a complete set of analytic expressions for the spatial OAM modal crosstalk  within the framework of perturbation theory. The obtained expressions include explicit formulas for the $2\pi$ walk-off length, and provide deeper insight into mode-mixing and thus the crosstalk behavior as a function of ellipticity and the topological charge carried by the OAM mode. The $2\pi$ walk-off length of a given input OAM mode refers to the distance over which the mode transforms itself into its degenerate partner characterized by a topological charge of opposite sign, and back into itself, under the continuous impact of the perturbation (ellipticity here). We further investigate the effect of  spin-orbit interaction \blk and provide approximate \blk bounds on the applicability of the the derived expressions. The developed theory is illustrated with application to  step-index  few mode and  \blk multimode fibers.
\section{Scalar Wave Equation for the Slightly Elliptical Fiber}
We first write the scalar wave equation for the propagation of an OAM mode through an unperturbed fiber as an eigenvalue equation
\begin{equation}
    HO_{l,m}(r,\theta)=\beta_{l,m}^2O_{l,m}(r,\theta),
\end{equation}
where the Hermitian operator $H= \vec{\nabla}_t^2+k^2n^2(r)$; $k=2\pi/\lambda$, where $\lambda$ is the wavelength; $\vec{\nabla}_t^2$ is the transverse Laplacian; $n^2(r)=n_1^2(1-2\Delta f(r))$; $f(r)$ is the index profile as a function of the radial cylindrical coordinate $r$ only, $\Delta= (n_1^2-n_2^2)/(2n_1^2)$ is the index profile height parameter, with $n_1>n_2$ \cite{snyder}; parameters $l$ and $m$ are respectively the topological charge (azimuthal mode number) and the radial mode number of the unperturbed OAM mode (denoted $OAM_{l,m}$) with an amplitude
\begin{equation}
O_{l,m} (r,\theta) =\frac{1}{\sqrt{N_{l,m}}}F_{l,m}(r) e^{il\theta},
\end{equation}
 where $N_{l,m}$ is the normalization constant; the eigenvalue $\beta_{l,m}^2$ is the square of the propagation constant of the $OAM_{l,m}$ mode, which gives the mode its $z$ dependence: 
\begin{equation}
\psi_{l,m}(r,\theta,z)=O_{l,m}e^{i\beta_{l,m} z}
\end{equation}
 ($z$ axis coincides with the fiber's axis). The amplitude, $O_{l,m}$, originates in the  reduction of  $HE_{l+1,m}, HE_{-l-1,m}, EH_{l-1,m}, EH_{-l+1,m} (|l|>1)$ to $O_{l,m}\epsilon_+, O_{-l,m}\epsilon_-,O_{l,m}\epsilon_-, O_{-l,m}\epsilon_+$, respectively in the weakly guiding approximation ($\Delta<<1$) \cite{bhandari,bhandari6}; $\epsilon_\pm=(1/\sqrt{2})(\hat{x}\pm i\hat{y}$) are the left (+) and right (-) circular polarizations corresponding to a photon spin $+1$ and $-1$, respectively (in units of $h/(2\pi)$); for the $l=1$ case, $HE_{2,m}, HE_{-2,m}, \frac{1}{\sqrt{2}}(TM_{0,m}+\blk i\blk TE_{0,m}), \frac{1}{\sqrt{2}}(TM_{0,m}-iTE_{0,m})$ reduce to the products: $O_{1,m}\epsilon_+, O_{-1,m}\epsilon_-, O_{1,m}\epsilon_-, O_{-1,m}\epsilon_+,$ respectively\cite{bhandari,ramachandran}. 
\begin{figure}[htbp]
\vspace{-2mm}
  \centering
  \includegraphics[width=10cm]{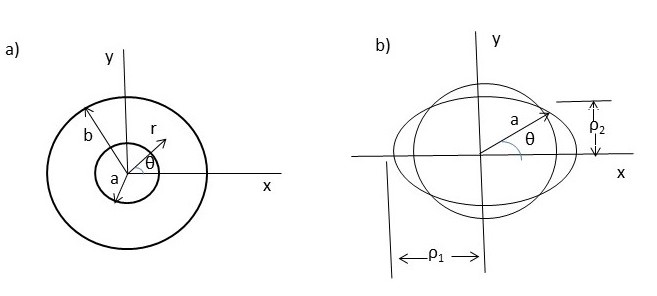}
\vspace{-3mm}
 \caption  {a)  Cross section of the multimode fiber with core radius $a$ and cladding radius $b (>>a); r$ and $ \theta$ along with the $z$ coordinate ($z$ axis coincident with the fiber axis) constitute the polar coordinates; the mode is assumed propagating in the $+z$ direction (out of the plane of the paper). b)  A slightly elliptical core shown relatively enlarged for clarity; semimajor axis $\rho_1=a(1+\epsilon)$ and semiminor axis $\rho_2=a(1-\epsilon)$ where ellipticity  $\epsilon=e^2/4 << 1$; $e$ is the eccentricity of the ellipse.}
\end{figure}
\\\\
We model the slightly elliptical fiber (see Fig. 1) \blk as a perfectly round fiber with a modified refractive index given by 
\begin{equation}
n'^{2}(r,\theta)=n^2(r) -2\epsilon\Delta n_1^2 (\partial{f(r)}/\partial{r})r\cos(2\theta);
\end{equation}
  ellipticity $\epsilon (<<1)$ is defined as the ratio of the difference of the semimajor axis and the semiminor axis to their sum, and equals $e^2/4$, where $e$ is the eccentricity of the ellipse \cite{snyder,alex}. 
Replacing $n$ with $n'$ in Eq. 1, the perturbed scalar wave equation is
\begin{equation}
(H+\epsilon\delta H)O_{l,m}'(r,\theta)=\beta_{l,m}'^2O_{l,m}'(r,\theta),
\end{equation}
where  $O'_{l,m}(r,\theta)$  and $\beta_{l,m}'$ are the corresponding perturbed mode amplitude and  propagation constant, respectively, and 
\begin{equation}
\delta H=-2k^2n_1^2\Delta\cos2\theta(\partial{f(r)}/\partial{r})r.
\end{equation}
The  scalar wave solution incorporating the z dependence is  written as $O'_{l,m}(r,\theta)e^{i\beta'_{l,m}z}$. The $O'_{l,m}$ amplitudes, like the $O_{l,m}$'s,  form a complete set. 

The effect of the perturbation $\epsilon\delta H$ is to cause mixing of an input OAM mode, $OAM_{l,m}$ with other (orthogonal) fiber OAM modes characterized by different pairs of parameters, $(l',m')\ne (l,m)$, \blk as described below.  \blk
\section{Perturbation Solution}

 Following \cite{bhandari2, bhandari}, we expand the perturbed amplitude $O_{l,m}'$ and the perturbed propagation constant $\beta_{l,m}'$ as
\begin{equation}
O_{l,m}'=O_{l,m}+\sum'_{n,k} a_{(l,m)(n,k)}^{(1)}O_{n,k}+\sum'_{n,k} a_{(l,m)(n,k)}^{(2)}O_{n,k}+....,
\end{equation}
\begin{equation}
\beta_{l,m}'^2=\beta_{l,m}^2+\beta_{l,m}^{2(1)}+\beta_{l,m}^{2(2)}+....,
\end{equation}
where the contributions in different orders of the perturbation parameter $\epsilon (<<1)$ are indicated by the superscripts in parentheses, and the prime on the summation implies $O_{n,k}\ne O_{l,m}$; all amplitudes are assumed normalized. 
\\\\
\blk
After inserting the above series in Eq. 5, taking the necessary scalar products to solve in different perturbation orders \cite{landau,mw, soliverez},  we obtain the analytic expressions  for the mixing coefficients and the propagation constant corrections:
\begin{equation}
 a_{(l,m)(n,k)}^{(1)}=\frac{\epsilon\delta H_{(n,k)(l,m)}}{(\beta_{l,m}^2-\beta_{n,k}^2)}, 
\end{equation}
\begin{equation}
 a_{(l,m)(n,k)}^{(2)}=\epsilon^2\bigg( \sum_{r,s}\frac{\delta H_{(n,k)(r,s)}\delta H_{(r,s)(l,m)}}{(\beta_{l,m}^2-\beta_{n,k}^2)(\beta_{l,m}^2-\beta_{r,s}^2)}-\frac{\delta H_{(n,k)(l,m)}\delta H_{(l,m)(l,m)}}{(\beta_{l,m}^2-\beta_{n,k}^2)^2}\bigg), 
\end{equation}
%
\begin{equation}
\beta^{2(1)}_{l,m}=\epsilon\delta H_{(l,m)(l,m)},
\end{equation}
\begin{equation}
\beta^{2(2)}_{l,m}=\epsilon^{2}\sum_{n,k}\frac{\delta H_{(l,m)(n,k)}\delta H_{(n,k)(l,m)}}{(\beta_{l,m}^2-\beta_{n,k}^2)}.
\end{equation}
 Eqs. 9-12 are standard equations up to second order in the perturbation parameter $\epsilon$ (higher order contributions, although more complicated in form, can similarly be determined, if needed). The matrix element
\blk
\begin{equation}
\begin{split}
\delta H_{(l,m)(n,k)}=&<O_{l,m}|\delta H|O_{n,k}>=\int_0^\infty\int_0^{2\pi} O_{l,m}^* (\delta H) O_{n,k} rdr d\theta\\
&= -2k^2n_1^2\Delta\int_0^\infty\int_0^{2\pi} O_{l,m}^* ((\partial{f(r)}/\partial{r})\cos(2\theta) O_{n,k} r^2dr d\theta;
\end{split}
\end{equation}
\blk the bra ($<$), ket ($>$) notation signifies a scalar product. This matrix element times $\epsilon$ represents the ellipticity-induced interaction (or coupling)  between the fields of the $OAM_{l,m}$ and $OAM_{n,k}$ modes. 
\blk Inserting Eq. 2 and integrating \blk over the azimuthal part, we immediately see that the $\delta H$ matrix elements are symmetric \blk $(\delta H_{(l,m)(n,k)}=\delta H_{(n,k)(l,m)}$), \blk and nonzero only when $|n-l|=2$, a selection  rule  also pointed out in [7]. Invoking  this selection rule, we
 \blk
 obtain simplified forms 
\blk
\begin{equation}
 a_{(l,m)(l\pm2,n)}^{(1)}=\frac{\epsilon\delta H_{(l\pm2,n)(l,m)}}{(\beta_{l,m}^2-\beta_{l\pm2,n}^2)}, 
\end{equation}
\begin{equation}
 a_{(l,m)(l\pm4,n)}^{(2)}=\epsilon^2 \sum_k\frac{\delta H_{(l\pm4,n)(l\pm2,k)}\delta H_{(l\pm2,k)(l,m)}}{(\beta_{l,m}^2-\beta_{l\pm2,k}^2)(\beta_{l,m}^2-\beta_{l\pm4,n}^2)}, 
\end{equation}
and so on.  The propagation constant corrections in \blk Eqs. 11 and 12 become \blk
\begin{equation}
\beta^{2(1)}_{l,m}= 0,
\end{equation}
\begin{equation}
\begin{split}
\beta^{2(2)}_{l,m}=&\epsilon^2\bigg(\sum_k\frac{\delta H_{(l,m)(l+2,k)}\delta H_{(l+2,k)(l,m)}}{(\beta_{l,m}^2-\beta_{l+2,k}^2)}\\
&+\sum_{k'}\frac{\delta H_{(l,m)(l-2,k')}\delta H_{(l-2,k')(l,m)}}{(\beta_{l,m}^2-\beta_{l-2,k'}^2)}\bigg).
\end{split}
\end{equation}
\\
\blk 
We  note that the mixing coefficient in first order perturbation, Eq. 14, not only depends upon $\epsilon\delta H_{(l\pm2,n)(l,m)}$, the ellipticity-induced interaction between $OAM_{l,m}$ and  $OAM_{l\pm2,n}$ modes, but also on what we term as a \emph{propagator}, which is the inverse of the difference between the squares of the propagation constant of the $OAM_{l,m}$ mode and the coupled  $OAM_{l\pm2,\blk n}$ mode; larger this  difference, smaller this coefficient. 
\begin{figure}[htbp]
\vspace{-2mm}
 \centering
 \includegraphics[width=10cm]{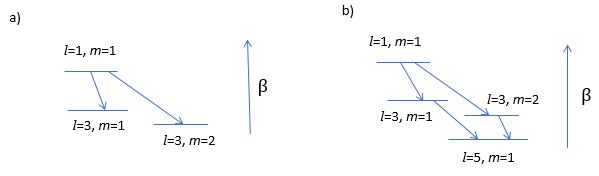}
\vspace{-3mm}
 \caption  {a) An input $OAM_{1,1}$ mode ($l=1,m=1$) couples to the fields of the neighboring $OAM_{3,1}$ and $OAM_{3,2}$ modes in first order perturbation; each arrow represents a coupling mediated by the ellipticity-induced interaction, $\epsilon\delta H$, times a propagator, which is the inverse of the difference of the propagation constant-squared  of the input mode and the connected mode, $OAM_{3,1}$ or $OAM_{3,2}$; the propagation constant increases from bottom to up, so $\beta_{1,1}>\beta_{3,1}>\beta_{3,2}$  b) Illustration of the second order perturbation effect, where input $OAM_{1,1}$ is  connected to $OAM_{5,1}$ through the intermediate $OAM_{3,1}$ mode as well as the intermediate $OAM_{3,2}$ mode, each sequence of connections making a separate contribution in Eq. 15 through the summation index $k$. Because two steps (transitions) are involved, the mixing process is of order $\epsilon^2$; note that, in general,  there can be more than two intermediate states (with the same $l=3$ value), depending upon the fiber; see text.}
\end{figure}
The second order coupling of the input mode $OAM_{l,m}$ to $OAM_{l+4,n}$ mode takes place in two steps by first coupling to $OAM_{l+2,k}$, which in turn couples to $OAM_{l+4,n}$; it is therefore of order $\epsilon^2$ (all these couplings follow the $|\Delta l|=2$ rule discussed above).  The expression, Eq. 15, has two matrix elements and two propagators within the summation. In general,  coefficients, $a_{(l,m)(l\pm2,n)}^{(i)}, i\ge 1$, will have $i ~ \delta H$ matrix elements connected by the $\Delta l=\pm 2$ rule, and accompanied by an equal number of propagators, along with an appropriate sum over all the intermediate states, as exemplified in Eq. 15. These coefficients are dimensionless because $\epsilon \delta H$ (Eq. 6) has the units of the square of the propagation constant, which cancel the units of the propagators, equal in number.
Fig. 2 is a pictorial representation  of  the results in Eqs. 14 and 15 for an $l=1,m=1$ mode. 
The first order correction to the propagation constant, Eq. 11, becomes zero due to the selection rule (see Eq. 16), while the second order correction, Eq. 12, is now given by Eq. 17; the number of of propagators is one less than the number of matrix elements in the numerator, consistent with the dimensionality of the square of  a propagation constant.  
\\\\
It is also important to note here that the selection rule $|\Delta l|=2$ also allows higher order transitions, which have been ignored in the above results. 
As an example, in first order perturbation, an input $OAM_{4,m}$ mode will couple to both  $OAM_{6,n}$ and $OAM_{2,n}$ modes ($n\ge 1)$ in a single step (Eq. 14). But perturbation theory (governed by the selection rule) also allows  the following types of couplings:  the input $OAM_{4,m}$ mode couples first to  to an $l=6$ mode ($OAM_{6,k}, k\ge1$), which then couples to $l=4 ~(OAM_{4,k'}, k'\ge 1)$, which in turn couples (transitions) back to the $OAM_{2,n}$ mode. There are three  steps involved, each step contributing a factor of the perturbation parameter, $\epsilon$; this mode of coupling is therefore weaker than the single step process (Eq. 14) by $\epsilon^2$ (recall $\epsilon<<1$). Such \emph{weaker} terms are ignored in arriving at Eq. 14. Weaker terms are also generated in the case of coupling of $OAM_{l,m}$  to $OAM_{l\pm4}$ in Eq. 15, but they are similarly ignored. 
\blk
\\\\
In what follows, OAM modes with negative topological charge will be indicated by placing an explicit minus sign. \blk  Now, $\beta_{-l,m}=\beta_{l,m}$, a consequence of degeneracy between $OAM_{-l,m}$ and $OAM_{l,m}$ modes; this follows from the $l^2$ dependence of the wave equation \cite{snyder}.  Furthermore, $\delta H_{(-l,m)(-l',n')}=\delta H_{(l,m)(l',n')}$, which is seen  from Eq. 13. \blk Thus, when $l$ is replaced by $-l$ in Eqs. 14-17, identical right-hand expressions result.  In other words, the perturbation series
\begin{equation}
O_{-l,m}'=O_{-l,m}+\sum'_{n,k} a_{(-l,m)(n,k)}^{(1)}O_{n,k}+\sum'_{n,k} a_{(-l,m)(n,k)}^{(2)}O_{n,k}+....,
\end{equation}
\blk
transforms to 
\begin{equation}
O_{-l,m}'=O_{-l,m}+\sum'_{n,k} a_{(l,m)(n,k)}^{(1)}O_{-n,k}+\sum'_{n,k} a_{(l,m)(n,k)}^{(2)}O_{-n,k}+....,
\end{equation}
where $a_{(l,m)(n,k)}^{(1)}$ and $a_{(l,m)(n,k)}^{(2)}$ are given by Eqs. 14 and 15, as in the $+l$ case. 
\blk
%
%
\subsection{Breaking of Degeneracy between $OAM_{l,m}$ and $OAM_{-l,m}$ Modes}
The modes $OAM_{l,m}$ and $OAM_{-l,m}$ are degenerate in the subspace spanned by these two (orthogonal) modes. However, because of the symmetry of the wave equation with respect to the parameters $l$ and $-l$, the diagonal elements of the $2~x~2$ matrix in the $(l,-l)$ subspace will be equal to each other, and similarly, the off-diagonal elements. Further, invoking the  $|\Delta l|=2$ selection rule, we immediately see  that the  off diagonal element that connects the two degenerate states \blk will become \blk nonzero in perturbation order  $l$ only; this is due to the fact that the selection rule requires any transition between them to occur in $l$ steps:  $l\rightarrow l-2 \rightarrow l-4 \rightarrow ....\rightarrow -l+4 \rightarrow -l+2 \rightarrow -l$; each step provides  a multiplying factor of $\epsilon$, resulting in a $\epsilon^l$ dependence of the off-diagonal element (see Fig. 3 as an illustration). 
\blk
\begin{figure}[htbp]
\vspace{-2mm}
  \centering
  \includegraphics[width=10cm]{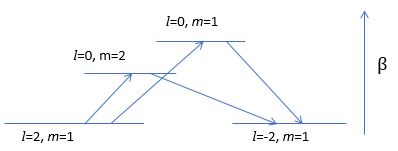}
\vspace{-3mm}
 \caption {Illustration of the transition from $l=2,m=1$ state to $l=-2,m=1$ state, which occurs in two steps due to the $|\Delta l|=2$ rule; each sequence of two transitions contributes separately in the summation of  Eq. 22 (see text).}
\end{figure}
\blk
 In effect, we have a symmetric matrix, \blk which needs to be diagonalized. A formal procedure in Appendix A yields the eigenamplitudes:
\begin{equation}
O^\pm_{l,m}=1/\sqrt{2}(O_{l,m}\pm O_{-l,m}),
\end{equation}
with corresponding eigenvalues 
\begin{equation}
(\beta'^{\pm}_{l,m})^2=\beta^2_{l,m}+\beta_{l,m}^{2(2)} \pm \gamma_{l,m}.
\end{equation}
 $\beta_{l,m}^{2(2)}$ is given by Eq. 17 and 
\blk
\begin{equation}
\begin{split}
\gamma_{l,m}=&(\epsilon)^{l}\sum\frac{\delta H_{(l,m)(l-2,n)}\delta H_{(l-2,n)( l-4,p)}....}
{(\beta_{l,m}^2-\beta_{l-2,n}^2)(\beta_{l,m}^2-\beta_{l-4,p}^2)....}\\
&\frac{....\delta H_{(-l+4,q)(-l+2,j)}\delta H_{(-l+2,j)(-l,m)}}{....(\beta_{l,m}^2-\beta_{-l+4,q}^2)(\beta_{l,m}^2-\beta_{-l+2,j}^2)}~ (l>1);
\end{split}	
\end{equation}
\begin{equation}
\gamma_{l,m}= \epsilon\delta H_{(l,m)(-l,m)}~(l=1).
\end{equation}
 Each $\delta H$ matrix element in the numerator of Eqs. 22 and 23 is a $|\Delta l|=2$ transition, there being $l$ in all; \blk the number of propagators is $l-1$. \blk The sum runs over all the radial mode solutions of the intermediate OAM states (repeated indices in Eq. 22).  Illustrating further,  for the $l=3, m=1$ case, $\gamma_{3,1}$ in Eq. 22 will have products of matrix elements $\delta H_{(3,1)(1,n)},\delta H_{(1,n)( -1,p)}, \delta H_{(-1,p)(-3,1)}$ and propagators $(\beta_{3,1}^2-\beta_{1,n}^2)^{-1}, (\beta_{3,1}^2-\beta_{-1,p}^2)^{-1}$,  along with the $\epsilon^3$ multiplying factor (repeated indices imply summation); this is a three-step transition. Note that $\beta_{-1,p}^2 =\beta_{1,p}^2$ and $\delta H_{(-1,p)(-3,1)}=\delta H_{(1,p)(3,1)}=\delta H_{(3,1)(1,p)}$ due to the symmetries discussed earlier. 
\\\\
\blk
\emph{Physical Interpretation}: \blk The linear combinations  given in Eq. 20 are the field amplitudes of the corresponding \emph{Linearly Polarized (LP)} modes, the $+$ sign corresponding to the $LP^{(a)}_{l,m}$ spatial mode and the $-$ sign to the $LP^{(b)}_{l,m}$ spatial mode, with intensities proportional to $\cos^2(l\theta)$ and $\sin^2(l\theta)$, respectively; they are, however,  now slightly nondegenerate (see Eq. 21). \blk If we further invert Eq. 20, we obtain
\begin{equation}
O_{\pm l,m}=1/\sqrt{2}(O^+_{l,m}\pm O^-_{l,m}).
\end{equation}
Consider now an $OAM_{l,m}$ mode, $\psi_{l,m}\blk(z)\blk=O_{l,m}e^{i\beta_{l,m}z}$ (see Eq. 3), incident on the slightly elliptical fiber at $z=0$. In view of the result, Eq. 24, this mode  will travel  down the fiber as an equal  superposition of two slightly nondegenerate $LP$ fields with propagation constants, $\beta^{'+}_{l,m}$ and $\beta^{'-}_{l,m}$, as given by Eq. 21. After traveling a distance $L$ within the fiber, its composition will change to
\begin{equation}
 \psi^{(e)}_{l,m}(L)= \frac{1}{\sqrt{2}}\Big(O_{l,m}^+e^{i\beta'^+_{l,m}L}+O^-_{l,m}e^{i\beta'^-_{l,m}L}\Big),
\end{equation}
where the superscript $e$ refers to the slight ellipticity of the fiber (we have ignored here the interaction with other modes, which will be done in the next section).  We now substitute Eqs. 20 and 21 into the above equation and  use the fact that $\beta^{'+}_{l,m}+\beta^{'-}_{l,m}\approx 2\beta_{l,m}$, since $ \beta^{(2)}_{l,m}<<\beta_{l,m}$. This yields after some algebra
\begin{equation}
\psi_{l,m}^{(e)}( L)=\Big(\cos(\pi L/L^{(2\pi)}_{l,m})O_{l,m}+ i \sin(\pi L/L^{(2\pi)}_{l,m}) O_{-l,m}\Big)e^{i\beta_{l,m}L},
\end{equation}
where
\begin{equation}
L^{(2\pi)}_{l,m}=\frac{2\pi}{\blk\Delta\beta'_{l,m}}\blk \approx \frac{2\pi\beta_{l,m}}{\gamma_{l,m}};
\end{equation}
 $\blk\Delta\beta'_{l,m}\blk=\beta'^+_{l,m}-\beta'^-_{l,m}\approx ((\beta^{'+}_{l,m})^2-(\beta^{'-}_{l,m})^2))/(2\beta_{l,m})=\gamma_{l,m}/\beta_{l,m}$ (see Eqs. 21). The entering $OAM_{l,m}$ mode oscillates into and out of its degenerate partner under the impact of  ellipticity; $L^{(2\pi)}_{l,m}$ is the $2\pi$ walk-off length, which is the distance over which a given mode transforms into its degenerate partner, and back into itself. From Eq. 26, we  see that at $L=L^{(2\pi)}_{l,m}/4$, the input $OAM_{l,m}$ has partly transformed  into its degenerate partner, $OAM_{-l,m}$, which now has the same amplitude as the parent mode, $OAM_{l,m}$.  At $L=L^{(2\pi)}_{l,m}/2$, the input mode has completely transformed into its degenerate partner (see Fig. 4). This cycle of interconversion of the two degenerate modes continues as $L$ increases further.
 \\\\
From Eqs. 27, 22, and 23, we also find that for given topological charge $l$, $L^{(2\pi)}_{l,m} \propto \epsilon^{-l}$. Thus,  larger the ellipticity $\epsilon$, smaller the $2\pi$ walk-off length. \emph{Physically, this implies that as the deviation from fiber circularity (ellipticity $\epsilon$) increases, an ellipticity-induced torque acting on the given $OAM_{l,m}$ mode increases, thus shortening the distance over which the $OAM_{l,m}$ mode converts into the $OAM_{-l,m}$ mode  as depicted in Fig. 4; this conversion, which occurs over half the $2\pi$ walk-off length, involves an OAM transfer  of $2l$ in magnitude (such a transfer of OAM also has been discussed in connection with fiber bends \cite{bhandari}). Similarly, for fixed $\epsilon$, as the required OAM transfer of $2l$  increases with $l$, we expect the transition to $OAM_{-l,m}$ mode to take place over a longer fiber length.} In other words, the $2\pi$ walk-off length $L^{(2\pi)}_{l,m}$ increases with $l$ for fixed $\epsilon$. These features of the $2\pi$ walk-off length are further illustrated numerically in Section 6 with reference to a step-index fiber.
\begin{figure}[htbp]
\vspace{-2mm}
  \centering
  \includegraphics[width=10cm]{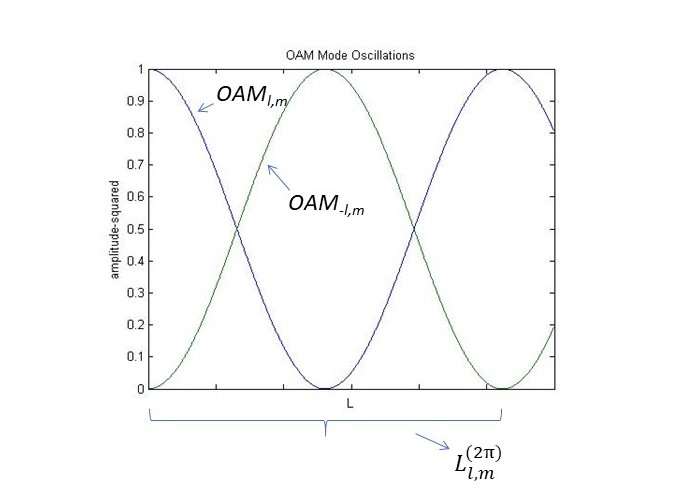}
\vspace{-3mm}
 \caption {Amplitude-squared, equal to $\cos^2(\pi L/L^{(2\pi)}_{l,m})$ and $\sin^2(\pi L/L^{(2\pi)}_{l,m})$ for the $OAM_{l,m}$ and $OAM_{-l,m}$ modes, respectively, as a function of $L$ (see Eq. 26); at $L=L^{(2\pi)}_{l,m}/2$, the input $OAM_{l,m}$ mode has transformed completely into its degenerate partner, the $OAM_{-l,m}$ mode.}
\end{figure}
\blk
\subsection{New Perturbation Series}
\blk
Perturbation fixes the appropriate linear combinations of the degenerate pairs of OAM modes  in accordance with Eq. 20 (see Appendix A). Therefore, we must work with these linear combinations, and replace the perturbation series in Eqs. 7 and 19 with the following series:
\begin{equation}
O'^{\pm}_{l,m}=O^{\pm}_{l,m}+\sum_{i=1,k} a^{(i)}_{(l,m)(n,k)}O^{\pm}_{n,k}
\end{equation}
derived  in Appendix A; $n=l\pm2i$, i.e., in the above sum, $n$ takes on both the values, $l+2i$ and $l-2i$, as in Eqs. 14 and 15 due to the $|\Delta l|=2$ rule.  The mode amplitudes  $O^{\pm}_{n,k}=1/\sqrt{2}(O_{n,k}\pm O_{-n,k})$ are linear combinations akin to $O^{\pm}_{l,m}$ (Eq. 20).  Like the latter, they represent the $LP$ fields corresponding to the $OAM_{n,k}$ mode; they are also slightly nondegenerate due to the slight fiber ellipticity, with $\blk(\blk\beta^{'+}_{n,k}\blk)^2\blk-\blk(\blk\beta^{'-}_{n,k}\blk)^2\blk=2\gamma_{n,k}$ (see Eqs. 22 and 23). Because of the linearity of the wave equation, $O'^{\pm}_{l,m}$ also satisfy the wave equation, Eq. 5, and form a complete set. 
\blk 
The $O^{'+}_{l,m}$ amplitudes expand in terms of the $O^{+}_{l',m'}$ amplitudes,  and similarly the  $O^{'-}_{l,m}$ amplitudes in terms of their corresponding counterparts, $O^{-}_{l',m'}$ (Eq. 28). Conversely, the $O^+_{l,m}$ amplitudes are expandable in terms of $O^{'+}_{l', m'}$ amplitudes  and similarly $O^-_{l,m}$ in terms of $O^{'-}_{l',m'}$. Hereafter,  the eigenmodes pertaining to amplitudes, $O^{\pm}_{l,m}$ will correspondingly be denoted by $OAM^{\pm}_{l,m}$. 

\blk
In the next section, we use the above series to determine the complete expression that includes the mixing of the $OAM_{l',m'}$ modes, $l'\ne -l$, not considered in Section 3.1.
\blk
\section{OAM Mode-Mixing and Crosstalk}
\blk For purposes of derivation, \blk consider an input $ OAM^+_{l,m}$  mode with amplitude $O^+_{l,m}$ entering a straight, slightly elliptical fiber. Let \blk$\psi\blk_{l,m}^{+(e)}\blk(r,\theta,z)|_{z=L}$ \blk denote the amplitude within the perturbed fiber at a distance $L$ from its entry point ($z=0$). Suppressing the arguments $r,\theta$ \blk(for brevity reasons)\blk,  we can then write
\begin{equation}
\blk\psi\blk_{l,m}^{+(e)}(L)\blk=\sum_{l',m'}{\eta^+_{(l,m)(l',m')}O^{'+}_{l',m'}}e^{i\beta^{'+}_{l',m'}L},
\end{equation}
\blk since the perturbed solutions $O^{'+}_{l',m'}$ also form a complete set (see Section 3.2). \blk
 $\eta^+_{(l,m)(l',m')}$ are the mixing coefficients to be determined. Using the boundary condition, \blk$\psi\blk_{l,m}^{+(e)}(0)\blk=O^+_{l,m}$, \blk and assuming orthonormality of $O^{'+}_{l,m}$'s, we immediately obtain
$\eta^{+}_{(l,m)(l',m')}= \int^\infty_0\int^{2\pi}_0 O^{*'+}_{l',m'}O^{+}_{l,m}r drd\theta$. Inserting Eq. 28, we get
\blk
\begin{equation}
\eta^{+}_{(l,m)(l',m')}= \delta_{ll'}\delta_{mm'}+\sum_{i=1}a^{(i)}_{(l',m')(l,m)}\delta_{l',l\pm 2i}.
\end{equation}
In the second term of Eq. 30, $l'$ is related to $l$ through the relationship: $l'=l\pm 2i$.
Substituting Eqs. 30 and 28 into Eq. 29 yields
\begin{equation}
\begin{split}
\psi^{+(e)}_{l,m}(L)=O^+_{l,m}e^{i\beta^{'+}_{l,m}L}&+\sum_{i=1,m'}a^{(i)}_{(l,m)(l',m')}O^+_{l',m'}e^{i\beta^{'+}_{l,m}L}
+\sum_{i=1,m'}a^{(i)}_{(l',m')(l,m)}O^+_{l',m'} e^{i\beta^{'+}_{l',m'}L}
\\
&+\sum_{i=1,m'}\sum_{i'=1,m''}a^{(i)}_{(l',m')(l,m)}a^{(i')}_{(l',m')(l'',m'')}O_{l'',m''}e^{i\beta^{'+}_{l',m'}L},
\end{split}
\end{equation}
where $l'=l\pm 2i$ and $l''=l'\pm 2i'$.
\blk
An identical expression results for an input $OAM^-_{l,m}$ mode, except that the $-$ superscript replaces the $+$ superscript everywhere. For an incident $OAM_{l,m}$ mode in which we are interested, we combine the two results:
\begin{equation}
\blk\psi\blk_{l,m}^{(e)}( L)\blk=\frac{1}{\sqrt{2}}\Big(\blk\psi\blk_{l,m}^{+(e)}(L)+\blk\psi\blk_{l,m}^{-(e)}( L)\Big),
\end{equation}
since $O_{l,m}=\frac{1}{\sqrt{2}}(O^+_{l,m}+O^-_{l,m}$) (see Eq. 24).
%
%
\subsection{Solution up to First Order}
\blk Considering $i=1$ only in Eq. 31 and discarding the fourth (quadratic) term on the RHS,  we obtain
\begin{equation}
\blk\psi\blk_{l,m}^{+(e)}( L)\blk=O^+_{l,m}e^{i\beta^{'+}_{l,m}L}+\sum_{l'=l\pm 2,m'}a^{(1)}_{(l,m)(l',m')}O^+_{l',m'}e^{i\beta^{'+}_{l,m}L}-\sum_{l'=l\pm 2,m'}a^{(1)}_{(l,m)(l',m')}O^{+}_{l',m'}e^{i\beta^{'+}_{l',m'}L},
\end{equation}
where we have used the fact that $a^{(1)}_{(l',m')(l,m)}=-a^{(1)}_{(l,m)(l',m')}$ (see Eq. 14); each summation includes only two terms:  $l'=l\pm 2$. 
By replacing the  + sign of the superscript everywhere in the above equation with the - sign, we obtain the expression for the incident mode $OAM^-_{l,m}$:
\blk
\begin{equation}
\psi\blk_{l,m}^{-(e)}( L)=O^-_{l,m}e^{i\beta^{'-}_{l,m}L}+\sum_{l'=l\pm 2,m'}a^{(1)}_{(l,m)(l',m')}O^-_{l',m'}e^{i\beta^{'-}_{l,m}L}-\sum_{l'=l\pm 2,m'}a^{(1)}_{(l,m)(l',m')}O^{-}_{l',m'}e^{i\beta^{'-}_{l',m'}L}.
\end{equation}
\blk
Combining the two expressions in accordance with Eq. 32, we obtain, up to first order \blk in $\epsilon$\blk,
\begin{equation}
\begin{split}
&\blk\psi\blk_{l,m}^{(e)}( L)\blk=\Big(\cos(\pi L/L^{(2\pi)}_{l,m})O_{l,m}+ i \sin(\pi L/L^{(2\pi)}_{l,m}) O_{-l,m}\Big)e^{i\beta_{l,m}L}\\
& +2i\sum_{l'=l\pm 2,m'} a_{(l,m)(l',m')}^{(1)}\Big(\cos(\pi L/L^{(2\pi)}_{l',m'})O_{l',m'}\\
&+ i \sin(\pi L/L^{(2\pi)}_{l',m'}) O_{-l',m'}\Big)(\sin(\beta_{l,m}-\beta_{l',m'})L/2) e^{i(\beta_{l,m}+\beta_{l',m'})L/2}.
\end{split}
\end{equation}
\blk
\emph{Physical Interpretation}:
The first two terms correspond to the mixing of the input $OAM_{l,m}$ mode with its degenerate partner, $O_{-l,m}$ mode, which was discussed in detail in Section 3.1. \blk
Mixing with the other modes $(l'=l\pm2)$ is captured in the second part of the right-hand-side of Eq. 35, where the mixed modes also occur in degenerate pairs and the OAM modes of a degenerate pair oscillate into each other. However, this mixing overall is suppressed by the first order perturbation mixing coefficient.  Physically, the incoming $OAM_{l,m}$ mode  couples first to the $OAM_{l\pm2\blk,m'\blk}$ mode due to the $\Delta l=\pm 2$ rule with an amplitude given by Eq. 14. But as this coupled $OAM_{l\pm2,m'}$ mode travels down the \blk slightly elliptical \blk fiber, it also continuously feels the impact of the ellipticity-induced torque, changing gradually  into its degenerate partner in an oscillatory manner, in accordance with its characteristic  $2\pi$ walk-off length (Eq. 27). \blk A multiplicative $sine$ term is a consequence of the  interference of the input $OAM_{l,m}$ mode and the coupled  $OAM_{l\pm2,m'}$ modes.
\blk
\\\\
For the input OAM modes defined by a $-l$ topological charge, we 
\blk replace Eq. 32 with $\psi\blk_{l,m}^{(e)}( L)\blk=\frac{1}{\sqrt{2}}(\blk\psi\blk_{l,m}^{+(e)}(L)-\blk\psi\blk_{l,m}^{-(e)}( L))$, which is  consistent with the requirement, $\blk\psi\blk_{-l,m}^{(e)}( 0)\blk=O_{-l,m}=\frac{1}{\sqrt{2}}(O^+_{l,m}-O^-_{l,m})$, the latter  following from Eq. 24. Substituting  Eqs. 33 and 34, we obtain the same final expression  as in Eq. 35, except that   $l$ is replaced with  $-l$ and $l'$ with $-l'$ everywhere, and  
\blk 
 amplitudes, $O_{l,m}$ and $O_{l',m'}$ are interchanged with their degenerate partners, $O_{-l,m}$ and $O_{-l',m'}$, respectively, and $l\pm2$ in the the summation on the RHS  is replaced with $-l\pm2$.
\\\\
\blk
\emph{Special Cases}
\\\\
1)  For the $|l|=2$ case, we can have $l'=0$ (see Eq. 35), which is a nondegenerate case. Therefore, $\delta\beta^{'\pm}_{0,m}=0$, which implies from Eq. 27, $ L^{(2\pi)}_{0,m'}=\infty$. Substituting in Eq. 35, we  correctly obtain $ \sin(\pi L/L^{(2\pi)}_{0,m'})=0$ and $\cos(\pi L/L^{(2\pi)}_{0,m'})=1$. 
\\\\
2) For  the $l=0$ incident mode (the nondegenerate case), the cosine and the sine terms in Eq. 35 involving the $O_{l,m}$ and the $O_{-l,m}$ modes, respectively, become one and zero, as discussed above.
\blk
The summation, however, now includes two terms, one corresponding to $l'=2$ and the other corresponding to $l'=-2$, yielding the expression, 
\begin{equation}
\blk\psi\blk^{(e)}_{0,m}(L)\blk=O_{0,m}e^{i\beta_{0,m}L}+2i a^{(1)}_{(0,m)(2,m')}e^{i\pi L/L^{(2\pi)}_{2,m'}}(O_{2,m'}+O_{-2,m'})\sin(\beta_{0,m}-\beta_{2,m'})e^{i(\beta_{0,m}+\beta_{2,m'})L/2}e^{i\beta_{2,m}L}. 
\end{equation}
The expression is symmetric in the coupled modes $OAM_{2,m'}$ and $OAM_{-2,m'}$. Physically, the input $OAM_{0,m}$ mode couples with the $l'=\pm 2$ modes with equal strength, each producing its own $cosine,~sine$ oscillatory behavior, which cancels out with the net effect  of  a  common (extra) phase factor governed by the $2\pi$ walk-off length. 
 
 %
 \subsection{Crosstalk}
 Crosstalk (or charge weight)\cite{yue, wang2, bhandari}, $\chi_{(l,m)(n,k)}$ (expressed in dB) for the various component $OAM_{n,k}$ modes of the output, $\blk\psi\blk^{(e)}_{l,m}(L)$\blk,  is given by
\begin{equation}
\chi_{(l,m)(n,k)}(L) = 10log_{10}|<O_{n,k}|\blk\psi\blk^{(e)}_{l,m}(L)\blk>|^2=10 log_{10}\Big|\int_0^\infty\int_0^{2\pi} O_{n,k}^*\blk\psi\blk^{(e)}_{l,m}( L)rdrd\theta\Big|^2.
\end{equation}
Upon substitution of Eq. 35, we find the crosstalk with the input mode's degenerate partner, $OAM_{-l,m}$, is 
\begin{equation}\
\chi_{(l,m)(-l,m)}(L)=10log_{10}\sin^2(\pi L/L^{(2\pi)}_{l,m}).
\end{equation}
%
%
 For the content of the original  $OAM_{l,m}$ mode remaining within the admixture, Eq. 35, we find
\begin{equation}
\chi_{(l,m)(l,m)}(L)=10log_{10}\cos^2(\pi L/L^{(2\pi)}_{l,m}).
\end{equation}
%
%
 The crosstalk  with the neighboring modes, $l'= l\pm 2 ~( l'>0)$, is from Eqs. 35 and 37
\begin{equation}
\chi_{(l,m)(l',m')} =10log_{10}\big(4|a^{(1)}_{(l,m)(l',m')}|^2\cos^2(\pi L/L^{(2\pi)}_{l',m'})\sin^2((\beta_{l,m}-\beta_{l',m'})L/2)\big),
\end{equation}
 and with the corresponding degenerate partners, $l'=-l\mp 2 ~( l'<0),$ is given by
\begin{equation}
 \chi_{(l,m)(l',m')}=10log_{10}\big(4|a^{(1)}_{(l,m)(l',m')}|^2\sin^2(\pi L/L^{(2\pi)}_{l',m'})\sin^2((\beta_{l,m}-\beta_{l',m'})L/2)\big).
\end{equation}
For the case of $l'=0$, which occurs when $| l|=2$, we simply set $ L^{(2\pi)}_{0,m'}=\infty$ in Eqs. 40 and 41 to obtain $ \chi_{(\pm 2, m)(0,m')}=10log_{10}\big(4|a^{(1)}_{(2,m)(0,m')}|^2\sin^2((\beta_{2,m}-\beta_{l0,m'})L/2)\big)$.
%
The maximum possible value from Eqs. 40 and 41  is 
\begin{equation}
\chi^{(max)}_{(l,m)(l',m')} \approx 10log_{10}(4|a^{(1)}_{(l,m)(l',m')}|^2); 
\end{equation}
this maximum corresponds essentially to setting the magnitude of the sinusoidal factors in Eqs. 40 and 41  to unity; these amplitudes vary rapidly with $L$ due to the small propagation constant differences in the argument of the second sinusoidal factor in each of Eqs. 40 and 41. 
For the $l=0$ (nondegenerate) input mode, the crosstalk with its neighbors, the $OAM_{2,m'}$ and $OAM_{-2,m'}$ modes, is given by $\chi_{(0,m)(\pm2,m')}=10log_{10}(4|a^{(1)}_{(0,m)(2,m')}|^2\sin^2((\beta_{0,m}-\beta_{2,m'})L/2))$ (see Eq. 36).
\section{Validity of the Scalar Theory}
The crosstalk equations in Section 4.2 along with the expressions for the $2\pi$ walk off length (Eqs. 27, 22, and 23) and the admixed neighbor mixing amplitude of Eq. 14 (in first order perturbation), comprise the fundamental set of equations derived on the basis of scalar perturbation theory. The weakly guiding approximation permitted us to write the vector mode solutions as products of the spatial modes and the polarization states, which then facilitated the use of scalar wave equation to study the impact of fiber  ellipticity on the spatial modes alone and their evolution  as a function of propagation distance. Inherent in the use of the scalar wave equation is the assumption of the smallness of spin-orbit ($SO$) interaction, which is then neglected.  
\\\\
 Here we  \blk briefly explore  the impact of the presence of $SO$ interaction.  The $SO$ induced correction  to the scalar propagation constant, denoted  $\delta\beta^{2(SO)}_{(l,m)}$,  is  of  the order of  $l\Delta/a^2$  \cite{snyder, volyar, bhandari4}, being of opposite sign for the $-l$ state as compared to the $l$ state. This then implies that the diagonal elements of the $2~x~2$ symmetric matrix in the $(l,-l)$ subspace (see Section 3.1 and Eq. A. 20), hitherto  equal, will tend to become unequal, the change for one diagonal element being of the opposite sign compared to the other. Consequently, this can  affect the accuracy of  the derived results. In order for our results to be as accurate as possible, we therefore impose the stringent condition that the splitting of the degenerate spatial modes, $2\gamma_{l,m}$ due to ellipticity $\epsilon$ (Eqs. 21-23) far exceeds the splitting due to the $SO$  interaction, i.e., the $(l,-l)$  degeneracy is still considered broken primarily through the elliptic effects. 
 \blk A simple analysis based on the requirement $\delta\beta^{2(SO)}_{(l,m)}<<\gamma_{l,m}$ as well as the sufficiency of the first order perturbation theory considered in our derivations (see Appendix B, \cite{bhandari5}),  yields, in an approximate manner, bounds on $\epsilon$  \blk
\begin{equation}
\Big(\frac{lV}{2k^2a^2 n_1^2}\Big)^{1/l}\Big(\frac{2}{V}\Big)<< \epsilon<<\Big(\frac{2}{V}\Big).
\end{equation}
\blk The lower bound, denoted  $\epsilon_{lo}$,   arises from the requirement of  $\delta\beta^{2(SO)}_{(l,m)}<<\gamma_{l,m}$, while the upper bound, denoted $\epsilon_u$, originates in the requirement that $|a^{(1)}_{(l,m)(l',m')}|<<1$ (the lower bound is less stringent than the upper bound, see Appendix B). \blk $\epsilon_{lo}$ increases with the topological charge $l$, implying a  diminishing region of accuracy, \blk $\epsilon_u-\epsilon_{lo}$; this is consistent with increase in  $SO$ interaction with $l$.  If  $V (=kan_1(2\Delta)^{1/2})$ were kept constant \blk (say, by decreasing $\Delta$ and increasing core radius $a$),  \blk transitioning to a fiber with a larger value of core radius  decreases the lower bound, $\blk\epsilon_{lo}\blk$, thus expanding the region, $\epsilon_u-\epsilon_{lo}$. In general, for large values of $V $, as in a multimode fiber, the lower bound decreases but the  upper bound becomes tighter. Thus, the accuracy range, \blk as \blk determined by these bounds, depends upon the fiber parameters and the value of the topological charge $l$.  \blk Conversely,  Eq. 43 can be a useful input in the design of  fibers, which suffer slight ellipticity and are governable by scalar perturbation theories due to the generally low effects of spin-orbit orbit interaction. \blk Note also that  if the  above bounds are not strictly adhered to,  the derived analytic expressions may suffer in numerical accuracy, but the qualitative behavior (for example, the rise in the $2\pi$ walk-off length with topological charge $l$ or its decrement with increasing ellipticity, $\epsilon$) will likely hold; this is  also supported by  physical reasoning (see Section 3.1). A detailed error analysis is beyond the scope of the current work.
%
\blk
 \section{Application to a Step-Index  Fiber}
 As an illustration, we now apply the derived expressions to a  step-index fiber, whose solutions are well known and well studied \cite{snyder}: 
\begin{align}
O_{l,m}(r,\theta)=\frac{1}{\sqrt{N_{l,m}}}J_l(p_{l,m}r)e^{il\theta} \quad for\ r\le a \nonumber \\
=\frac{1}{\sqrt{N_{l,m}}}\frac{J_l(p_{l,m}a)}{K_l(q_{l,m}a)}K_l(q_{l,m}r)e^{il\theta} \quad for \ r\ge a,
\end{align}
where $N_{l,m}$, the normalization constant, can be analytically determined \cite{mw}; $J_{l}$ and $K_l$ are the Bessel and the modified Bessel functions, respectively; $p_{l,m}=\sqrt{k^2 n_1^2-\beta_{l,m}^2}$ and $q_{l,m}=\sqrt{\beta_{l,m}^2-k^2 n_2^2}$. The index profile $f(r)$ is a step-function equal to zero for $r\le a$ and equal to 1 for $r>a$. As a result, $\partial{f(r)}/\partial {r} = \delta(r-a)$. The matrix element for the $\delta H$ operator defined in Eq. 13 is now given by 
\begin{equation}
    \delta H_{(l,m)(l\pm2,m')}=-(k^2n_1^2\Delta)\frac{J_l(p_{l,m}a)J_{l\pm2}(p_{l\pm2,m'}a)}{\sqrt{N'_{l,m}}\sqrt{N'_{l\pm2,m'}}},
\end{equation}
where   $N'_{n,k}= N_{n,k}/2\pi a^2$ is dimensionless.  Propagation constants required in the calculations are determined using the well-known characteristic equation for the scalar modes of  a step-index fiber\cite{snyder, buck}. All numerical calculations are done in MatLab.

\subsection{Few Mode Fiber}
We assume $a=10 \mu m,~ n_1=1.45205, ~n_2=1.44681$. The index height profile  parameter $\Delta = (n_1^2-n_2^2)/(2n_1^2)=0.0036<<1$. For a wavelength, $\lambda=1.55 \mu m$,  normalized frequency $V=2\pi a (n_1^2-n_2^2)^{1/2}/\lambda=4.996$. As a result, only  the $OAM_{0,1}, OAM_{1,1}, OAM_{-1,1}, OAM_{2,1}, OAM_{-2,1}$ and $OAM_{0,2}$ modes are supported. From Eq. 23 and Eq. 22, 
\begin{equation}
\gamma_{1,1}=\epsilon\delta H_{(1,1)(1,-1)}
\end{equation}
and 
\begin{equation}
\gamma_{2,1}=\epsilon^2\Big (\frac{(\delta H_{(2,1)(0,1)})^2}{\beta^2_{2,1}-\beta^2_{0,1}}+\frac{(\delta H_{(2,1)(0,2)})^2}{\beta^2_{2,1}-\beta^2_{0,2}}\Big ).
\end{equation}
Now using Eqs. 27, 22, and 23,  we can calculate  the $2\pi$ walk-off lengths, which are displayed in Table 1 for various values of ellipticity. 
\begin{table}
\centering
\caption{The $2\pi$ walk-off length, $L^{(2\pi)}_{l,m}$, specified in meters, as a function of ellipticity $\epsilon$ for different input modes, $OAM_{l,m}$; it varies as $\epsilon^{-l}$ for fixed $l,m$ (see Eqs. 27, 22, and 23); normalized frequency $V=4.996$.} 
\begin{tabular}{ |c|c|c|c|c|c|} 
 \hline
$ l,m$&$ \epsilon=0.005$ & $\epsilon=0.010$&$\epsilon=\blk 0\blk .015$&$\epsilon=0.025$&$\epsilon=0.040$\\ 
\hline
 1,1 & 0.372& 0.186&0.124&0.074&0.047 \\ 
\hline
 2,1&29.\blk 18\blk & 7.3\blk 0\blk &\blk 3.24\blk &1.17&0.456\\ 
 \hline
\end{tabular}
\end{table}
Larger the ellipticity $\epsilon$, smaller the $2\pi$ walk-off length $L^{(2\pi)}_{l,m}$ due to the $\epsilon^{-l}$ dependence. Likewise, larger the value of topological charge $l$, for the same ellipticity $\epsilon$ value, larger the $2\pi$ walk-off length (Section 3.1). This is what we observe in Table 1. \blk Note also that the ellipticity values for the $l=1$ case satisfy the  bounds of Eq. 43 ($\epsilon_{lo}=0.0003$ and $\epsilon_u=0.40$). For the $l=2$ case, $\epsilon_{lo}=0.015$, implying that the results in the table below this threshold may not be as accurate. 
%
\blk
\subsubsection{Mixing with Neighboring Modes}
\underline{Input: $OAM_{1,1}$}
\\\\
The $OAM_{1,1}$ input mode can only mix with $OAM_{-1,1}$ due to the $\Delta l=\pm 2$ rule (already considered above via Eq. 46 in connection with the $2\pi$ walk-off length). Referring to Eq. 35, excluding $l'=-1$, the only other allowed value of  $l'$ is 3, which is not supported by the fiber. So there is no other mode the input mode $OAM_{1,1}$ can mix with.
\\\\
\underline{Input: $OAM_{2,1}$}
\\\\
Besides undergoing the transformation into the degenerate partner, the $OAM_{-2,1}$ mode, the $OAM_{2,1}$ mode also mixes with the modes $OAM_{0,1}$ and $OAM_{0,2}$ with the amplitudes given by (see Eq. 35)

1) $|<O_{0,1}|\blk\psi\blk^{(e)}_{2,1}>|=2|a^{(1)}_{(2,1)(0,1)}\sin((\beta_{2,1}-\beta_{0,1})L/2)|=0.246\epsilon |\sin(5735L)|$

2) $|<O_{0,2}|\blk\psi\blk^{(e)}_{2,1}>|=2|a^{(1)}_{(2,1)(0,2)}\sin((\beta_{2,1}-\beta_{0,2})L/2)|=3.12\epsilon |\sin(915L)|$
\\\\
The numerical coefficient of the amplitude is larger in 2) than in 1) due to the closer proximity of the pair of modes $OAM_{2,1}, OAM_{0,2}$ in their propagation constant values as compared to the mode pair: $OAM_{2,1}, OAM_{0,1}$; these propagation constant differences are reflected in the numerical part of the argument of the sine functions. The amplitudes vary rapidly with $L$, but the maximum  amplitude corresponds to setting the sine term above to unity. These expressions are used in the crosstalk calculation in Section 6.1.2.
\begin{table}
\centering
\caption{ Crosstalk, $\chi_{(2,1)(l',m')}$ (in dB) for the various component $OAM_{l',m'}$ modes within the $OAM_{2,1}$ output mode mixture, $\blk\psi\blk^{(e)}_{2,1}$ as a function of $L$ (see Section 4.2); the ellipticity $\epsilon$ is fixed at 0.025. The $2\pi$ walk-off length $L_{2,1}^{(2\pi)} = 1.17m$ (see Table 1).}
\begin{tabular}{ |c|c|c|c|c|c|} 
 \hline
$ l',m'$& L=0.10m & L=0.25m & L=0.50m & L=1.00m & L=1.50m\\ 
\hline
 2,1 & -0.32&-2.12&-12.91&-0.94&-3.98\\ 
\hline
 -2,1&-11.53& -4.12&-0.23& -7.12& -2.22\\
 \hline
 0,1&-44.33&-44.89&-47.33& -44.22 & -46.95\\ 
 \hline
 0,2&-30.48&-27.30&-22.87& -25.07& -30.86\\ 
 \hline
\end{tabular}
\end{table}
\subsubsection{Crosstalk}
For $\epsilon =0.025$,  we show in Table 2 crosstalk calculated in accordance with Eqs. 37-40.  At $L=0.10m$, we see that the output mixture is practically all $OAM_{2,1}$ mode with little $OAM_{-2,1}$ mode content. However, at $L=0.25m$, roughly one quarter of the $2\pi$ walk-off length value, the contents of the $OAM_{2,1}$ and its degenerate partner, $OAM_{-2,1}$ mode are approximately the same, with the latter becoming the major constituent of the output mixture at $L=0.50m$ (roughly half the $2\pi$ walk-off length). At $L=1m$, slightly less than the $2\pi$ walk-off length, the   $OAM_{2,1}$ mode is the main contributor to the mixture, and  this cycle of mode conversion from the $OAM_{2,1}$ mode into $OAM_{-2,1}$ mode, and back into the original $OAM_{2,1}$ mode, continues as $L$ is increased beyond the $2\pi$ walk-off length.  The contents of the other admixed modes $l'\ne \pm l$ remain very low being (upper) bounded by values of $-44.22dB$ and $-22.16dB$ (calculated from Eq. 42) for $\chi^{(max)}_{(2,1)(0,1)}$ and $\chi^{(max)}_{(2,1)(0,2)}$, respectively.
\subsubsection{OAM Mode Propagation in a Multiplexed Environment}
 If we now adopt a  crosstalk criterion of $\chi_{(2,1)(l',m')} \le -20 dB$ for successful $OAM_{2,1}$ transmission, then  the maximum  propagation distance, denoted $L^{(max)}_{2,1}$, that is allowed is given by setting $10log_{10}(\sin^2(\pi L^{(max)}_{2,1}/L^{(2\pi)}_{2,1})\blk)\blk=-20$, which is needed to suppress the admixed $-l$ degenerate $OAM$ component (see Eq. 38). This gives rise to multiple solutions due to the multi-valued $\sin^{-1}$ function:  $L^{(max)}_{l,1}=0.037 m,~ L^{(max)}_{l,1}=(\pm 0.037 +n)L^{(2\pi)}_{l,1} (m)$, where $n>0$ is an integer. In other words, the propagation distance $L$ must satisfy the constraint $L\le L^{(max)}_{2,1}=0.037 m, (n L^{(2\pi)}_{2,1}-0.037m)<L \le ( n L^{(2\pi)}_{2,1}+0.037m)$,   where $n (>0)$ is an integer. Because $L^{(2\pi)}_{2,1}=1.17m$, this implies permitted transmissions at most of the order of a  few meters suitable  for inter-shelf/rack distances within data centers. Furthermore, the $2\pi$ walk-off length of the degenerate partner is identical and the nondegenerate modes, $OAM_{0,1}$ and $OAM_{0,2}$, which can only  mix with the $OAM_{2,1}$ and the $OAM_{-2,1}$ modes due to the $\Delta l=\pm 2$ selection rule,  satisfy the suppression constraint of $-20dB$ (see Table 2). Consequently, under this criterion we can have simultaneous transmission of the four spatial modes, $OAM_{2,1}, OAM_{-2,1}, OAM_{0,1},$ and $OAM_{0,2}$ in a multiplexed environment, thereby increasing the data transmission capacity four-fold.  Additional tables can be constructed for different sets of ellipticity and fiber parameters to gain insight into the constraints imposed by ellipticity in the design of such fibers. 
\blk
\subsection{Multimode Fiber}
\blk 
\begin{footnotesize}
\begin{figure}[htbp]
  \centering
  \begin{subfigure}[b]{0.45\linewidth}
    \includegraphics[width=\linewidth]{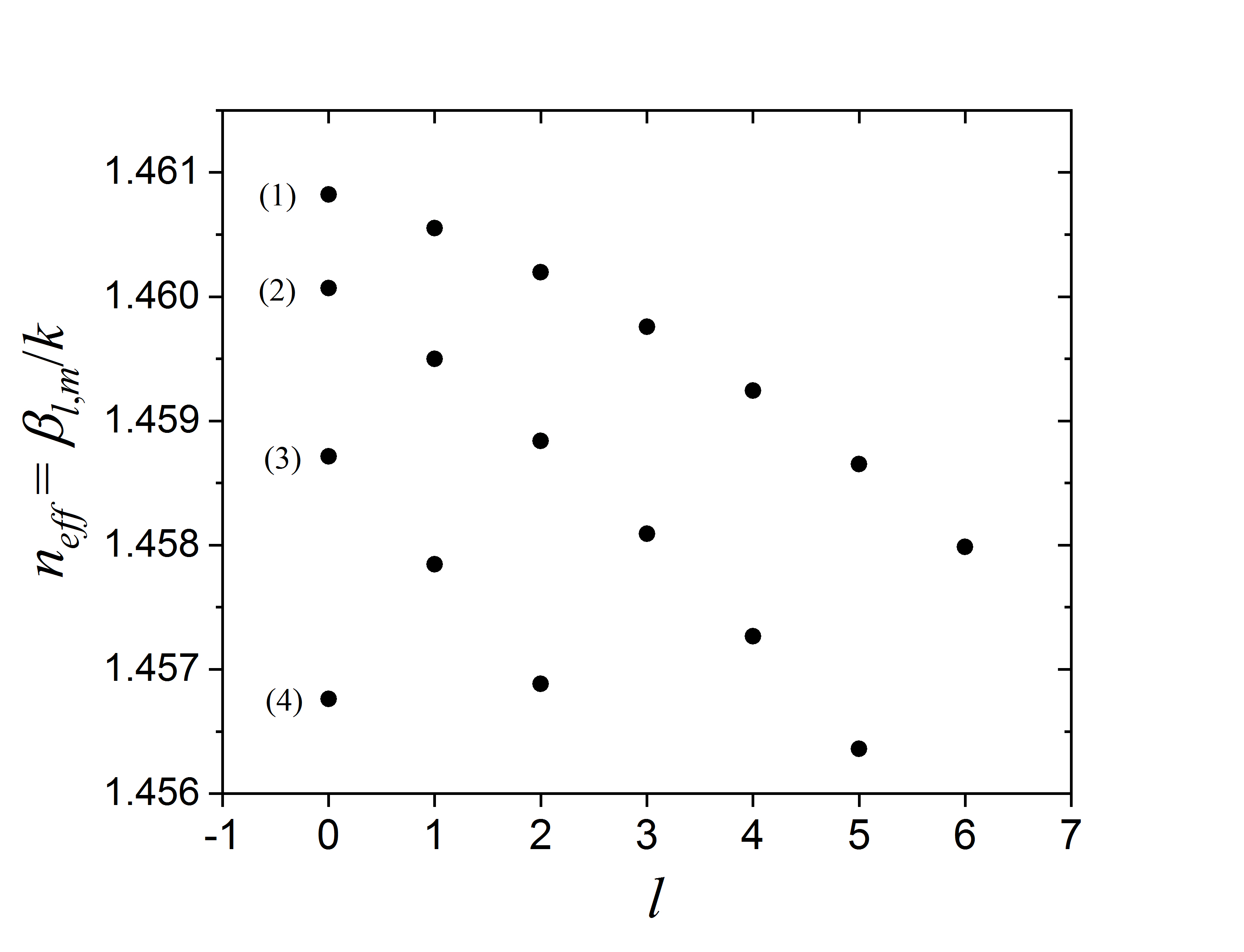}
  \end{subfigure}
 \begin{subfigure}[b]{0.45\linewidth}
    \includegraphics[width=\linewidth]{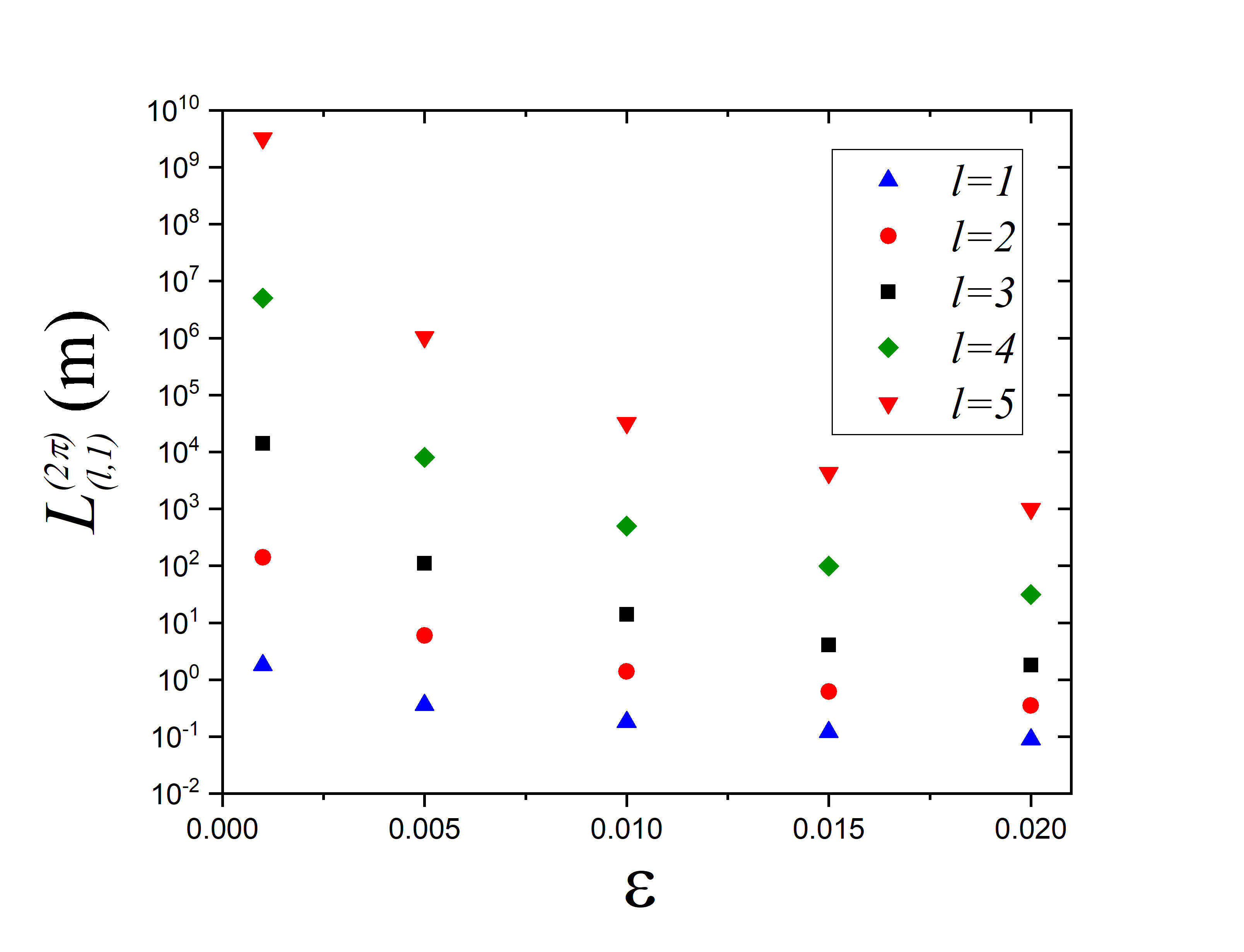}
  \end{subfigure}
 \begin{subfigure}[b]{0.45\linewidth}
    \includegraphics[width=\linewidth]{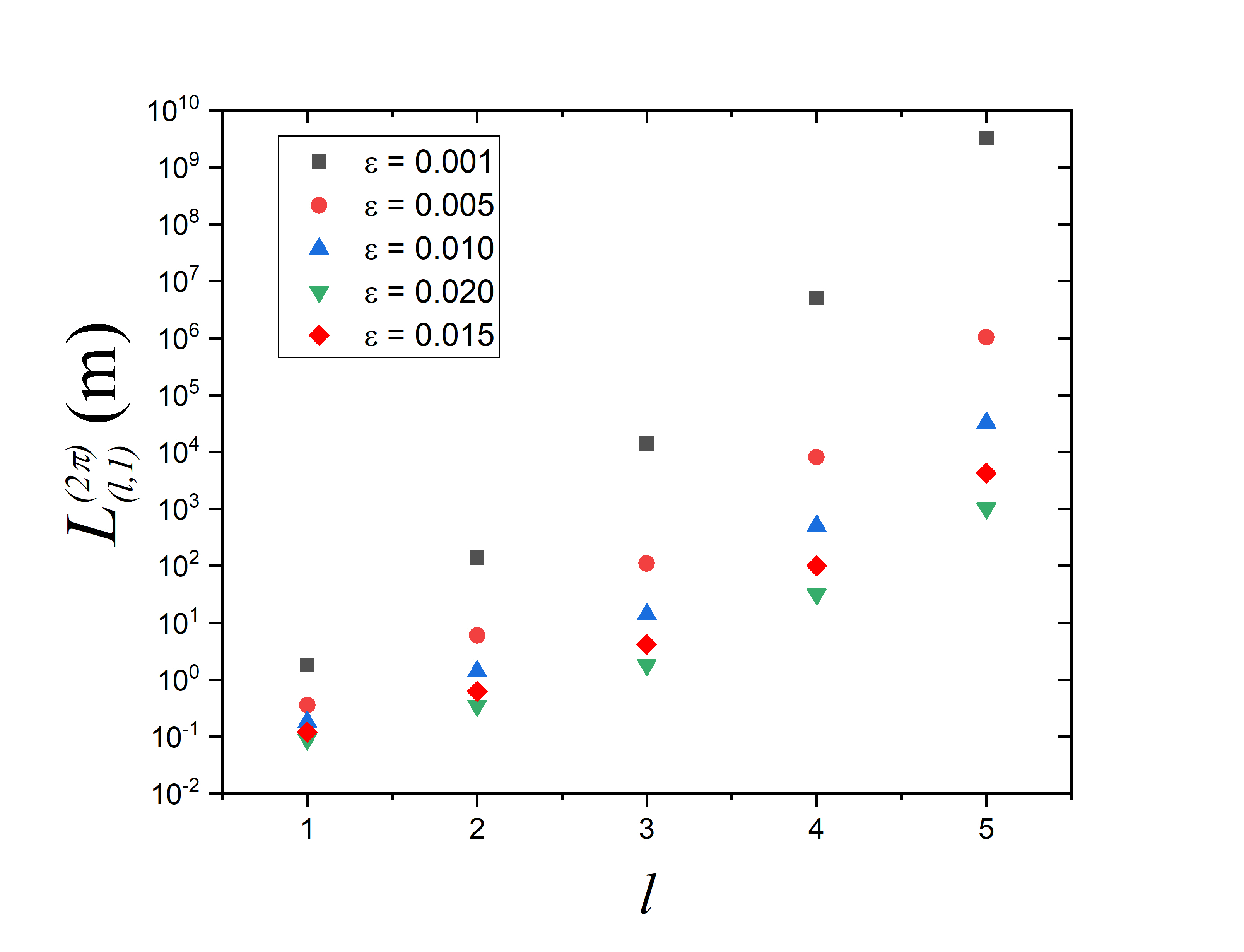}
  \end{subfigure}
\begin{subfigure}[b]{0.45\linewidth}
    \includegraphics[width=\linewidth]{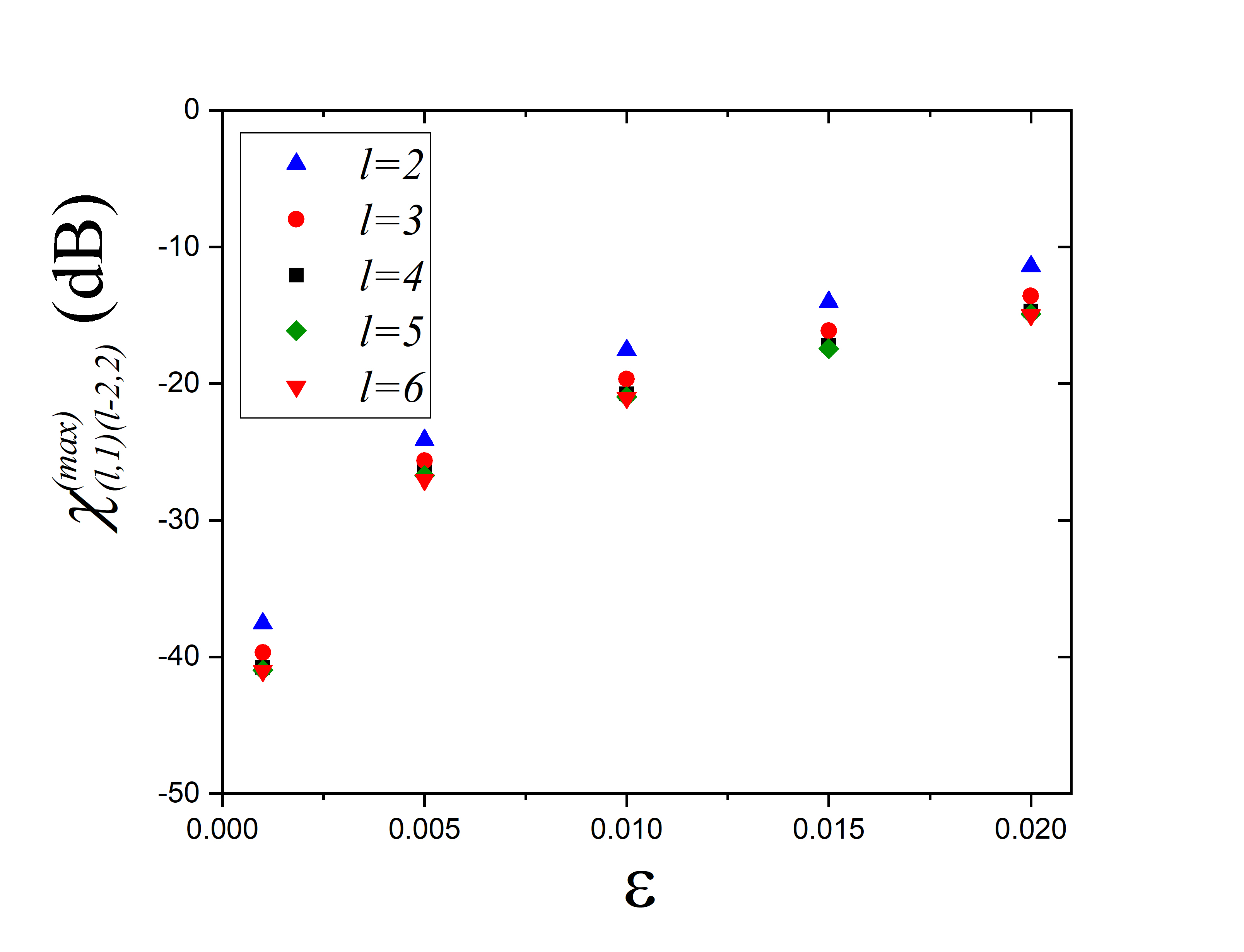}
  \end{subfigure}
  \label{multimode}
\caption  {a) The effective refractive indices, $n_{eff}$'s  shown with a cut-off of 1.456, for most modes up to $l=6$;  the values decrease with increasing $l$ and $m$, which is shown explicitly in parentheses for the $l=0$ mode  b) The $2\pi$ walk-off length, $L^{(2\pi)}_{(l,1)}$, as a function of ellipticity $\epsilon$ for different values of $l$; for fixed $l$, $L^{(2\pi)}_{(l,1)}$ varies inversely as $\epsilon^l$  c)   The $2\pi$ walk-off length, $L^{(2\pi)}_{(l,1)}$, as a function of topological charge $l$ for different values of $\epsilon$ \blk d) $ \chi_{(l,1)(l-2,2)}^{(max)}$, the maximum possible crosstalk of $OAM_{l,1}$ mode with the neighboring mode, $OAM_{l-2,2}$,   as a function of $\epsilon$ for different $l$ values; see text.}
\end{figure}
\end{footnotesize}

We assume  $a=25 \mu m, n_1=1.461$, $n_2=1.444$; these parameters correspond to a real \emph{ThorLabs} multimode fiber. $\Delta= (n_1^2-n_2^2)/(2n_1^2)=0.012$. We use $\lambda=1.55\mu m$.  The normalized frequency $V$ is 22.3. Although the maximum allowed value of $l$ is 18, we consider $l$ up to 6. Fig. 5a shows the effective refractive indices for the various OAM modes. The values decrease with increasing $l$ and $m$.  For fixed $l$, the spacing between consecutive $m$ indices also increases, the increase being larger for larger $l$ values. The effective refractive index difference, $\Delta n_{eff}$, is the smallest for the $OAM_{5,1}$ and the $OAM_{0,3}$ mode pair and equals $0.64~ x ~10^{-4}$. However, the $\Delta l= \pm 2$ rule prevents any transition between these two modes in any order of perturbation theory. The $OAM_{2,1}$ and the $OAM_{0,2}$ modes, which are very close in proximity ($\Delta n_{eff}=1.25 ~x ~10^{-4}$) but also satisfy the $\Delta l=\pm 2$ rule, will mix in first order; the mixing coefficient, $a^{(1)}_{(2,1)(0,2)}$, is likely to be larger as compared with the mixing coefficient of the modes $OAM_{2,1}$ and $OAM_{0,1}$ that also satisfy the $\Delta l=\pm 2$ transition rule, but have a much larger value of $\Delta n_{eff}= 6.26 ~x ~10^{-4}$. \blk In general, the proximity of the $OAM_{l,1}$ and the $OAM_{l-2,2}$ modes in their $n_{eff}$ values (see Fig. 5a) causes the latter to be the predominantly admixed neighboring mode. \blk
\\\\
 Fig. 5b shows  $L^{(2\pi)}_{(l,1)}$ as a function of ellipticity $\epsilon$ for different values of topological charge $l$. It is calculated using Eqs. 27, 22, and 23. It decreases as ellipticity $\epsilon$ is decreased, the drop being sharper for higher $l$ values; this is due to the $\epsilon^{-l}$ dependence. We note here that in the limit $\epsilon \rightarrow 0$, the  $2\pi$ walk-off length approaches infinity, consistent with the absence of crosstalk (see Eqs. 38, 41 and 14) for a perfectly round fiber ($\epsilon=0)$. In Fig. 5c, the ellipticity $\epsilon$ is fixed. The $2\pi$ walk-off length rises with $l$, the rise being sharper with $l$, which implies that the higher $l$ value modes will have relatively less crosstalk with their degenerate partners (see Eq. 38).
\blk
\begin{table}
\centering
\caption{The lower bound, $\epsilon_{l\blk o\blk}$ and the upped bound $\epsilon_u$ for different values of $l;~ \epsilon_u$ is independent of $l$ (see Eq. 43).} 
\begin{tabular}{ |c|c|c|} 
 \hline
$ l,m$&$ \epsilon_{lo}$ & $\epsilon_u$\\ 
\hline
 1,1 & $4.6X10^{-5}$& 0.09\\ 
\hline
 2,1&0.003&0.09\\ 
 \hline
3,1&0.010&0.09\\
\hline
4,1&0.019&0.09\\
\hline
5,1&0.027&0.09\\
\hline
\end{tabular}
\end{table}
\\\\
With respect to the impact of SO interaction on our results, we find that the  larger radius $a$ of the core of the multimode fiber compared to that of the few  mode fiber causes the $SO$ interaction to reduce; on the other hand, the elliptic effects leading to splitting of the degenerate modes are enhanced due to the relatively smaller  propagation constant differences on account of a larger value of $V$.  The  preferred region of $\epsilon$ (Eq. 43) expands to accommodate lower values of $\epsilon$, but diminishes as $l$ increases; \blk see Table 3, where the lower bound $\epsilon_{lo}$ increases with $l$, consistent with a rising value of the "undesired" SO interaction (of order $l\Delta/a^2)$; the desired ellipticity range, $\epsilon_u-\epsilon_{lo}$, therefore, reduces with $l$. If we were to increase the core radius  to $62.5 \mu m$ corresponding to another commercially available fiber, according to Eq. 43 and the fact that $V=ka\blk n_1 (2\blk\Delta\blk )\blk^{1/2}$, the lower bound $\epsilon_{lo}$ would decrease, for fixed $l$, by a factor $(50/62.5)^{1/l}(50/62.5)$;  the reduced $\epsilon_{lo}$ values then would be 0.074, 0.014, 0.020 for $l=3,4,5$, respectively. Thus, by increasing the core radius, we can accommodate lower values  of $\epsilon$ at higher values of $l$. Such analyses can be useful in the design of fibers with low spin-orbit interaction. \blk
%
\\\\
 In Table 4, we show the crosstalk experienced by the input $OAM_{3,1}$ and $OAM_{2,1}$ modes in a fiber with ellipticity $\epsilon=0.015,$  (which lies within the bounds defined by Table 3). While these modes change into their respective degenerate partners, they also mix with the neighboring modes, $OAM_{l\pm2,m'}$ with amplitudes proportional to   $a^{(1)}_{(\blk l\blk,m)(\blk l\blk\pm2,m')}$ determined from Eq. 14. 
The   crosstalk values displayed  in Table 4 are all seen to be below the maximum possible values of $\chi^{(max)}_{(3,1)(1,2)} =-16.14 dB$ and $\chi^{(max)}_{(2,1)(0,2)}=-14.02 dB$ (Eq. 42).
%
%
%
The $2\pi$ walk-off length for the $l=3$ modes is $4.15 m$. We note from the table that at $L=0.5m$, the $OAM_{3,1}$ mode is essentially pure with very little $OAM_{-3,1}$ content. However, at $L=1.0 m$, which is approximately one-quarter the $2\pi$ walk-off length, the contents of the  $OAM_{3,1}$ mode and its degenerate partner are roughly equal. At $L=2.0 m$ (approximately half the $2\pi$ walk-off length), the $OAM_{3,1}$ is basically converted fully into the  $OAM_{-3,1}$ mode. Similarly, at $L=4m$ (slightly less than the $2\pi$ walk-off length), most of the $OAM_{-3,1}$ mode has changed back into the $OAM_{3,1}$ mode. This oscillatory behavior continues, with a period equal to $4.15m$,  the $2\pi$ walk-off length. For the $\blk l\blk=2$ case, the $2\pi$ walk-off length is $0.63 m$; the changes here are more rapid on account of the smaller value of the $2\pi$ walk-off length. 
\begin{table}
\centering
\caption{Crosstalk, $\chi_{(l,m)(l',m')}$ (in dB) (see Section 4.2); $m=1$, $l=2$ and $3$. Ellipticity  $\epsilon$ is fixed at $0.015$. $L$ is the length of the fiber traversed by the $OAM_{l,1}$ mode input at one end of the fiber; $l'=l-2$.} 
\begin{tabular}{ |c|c|c|c|c|c|} 
 \hline
$ l',m'$&L=0.5m & L=1.0m & L=2.0m &L=4.0m & L=5.0m \\ 
\hline
 3,1 &-0.64&-2.77&-24.92&-0.06&-1.94\\ 
\hline
 -3,1&-8.65&-3.26&-0.01&-18.91 &-4.44\\ 
 \hline
1,2&-20.59&-23.07&-19.11&-34.04&-18.65\\
\hline
-1,2&-16.18&-17.55&-15.75&-25.99&-17.28\\
\hline
 2,1&-1.97&-11.35&-1.38&-6.82&-0.17\\
 \hline
 -2,1&-4.38&-0.33&-5.66&-1.01&-14.06\\
 \hline
0,2&-17.86&-16.69&-19.92&-16.26&-30.70\\
\hline
\end{tabular}
\end{table}
%
%
\subsection{Propagation distance in an OAM multiplexed environment}
\blk
If we now  consider a fiber which is perfect, except for ellipticity, and adopt a somewhat relaxed criterion of crosstalk $< -14dB$ with any admixed mode, then this criterion translates into a  requirement of  a maximum permitted propagation distance, $L^{(max)}_{l,1}$, given by $ 10log_{10}sin^2(\pi L^{(max)}_{l,1}/L^{(2\pi)}_{l,1})=-14$; this implies multiple solutions (as in Section 6.1.3) and leads to the constraint that the propagation distance $L$ must satisfy:
 $L \le 0.064 L^{(2\pi)}_{l,1},~ (n-0.064)L^{(2\pi)}_{l,1}<L<(n+0.064)L^{(2\pi)}_{l,1}, ~ l=2,3$. For example, L=4.40m satisfies the inequality for both $l=2$ and $l=3~ (n=1, 7$ ~for~$ l=3, 2$, respectively). This  has the implication that we can multiplex $OAM_{2,1},OAM_{-2,1}, OAM_{3,1}, OAM_{-3,1}$ and transmit without crosstalk exceeding $-14dB$ when $L=4.40 m$; this corresponds to  an inter-shelf/rack distance in a data center.

\section{Further Remarks}
%
\blk
\blk
1) Because the weakly guiding approximation causes the vector modes to be expressed as products of the spatial OAM mode considered here and the circular polarization (right or left) (Section 2),  the impact of fiber ellipticity is then the product of the impact of ellipticity on the spatial mode and the polarization mode, evaluated separately. In general, the initial  circular polarization  would become a linear combination of the left and right circular polarizations; the overall  final state (including polarizations) would be the product of the final spatial mode state as determined by our analysis and the final polarization state not determined here; this would, for example, result in a linear combination of the quartets: $O_{l,m}\epsilon_+, O_{-l,m}\epsilon_-,O_{l,m}\epsilon_-, O_{-l,m}\epsilon_+$, translating, respectively  to the vector modes: $HE_{l+1,m}, HE_{-l-1,m}, EH_{l-1,m}, EH_{-l+1,m} (|l|>1)$. Our intent here is to treat the impact of ellipticity on the spatial mode only, meaning we are not concerned with the final polarization state of the output modes. In the experimental domain, this corresponds to separating out the spatial modes, as is done, for example,  in\cite{huang, zhou} using a mode sorter, and measuring their individual intensities, without any regard to their polarization state. 
\\\\
2) The solutions, $O^{'\pm}_{l,m}$, Eq. 28, are the eigenmodes of the scalar wave equation: $(H+\epsilon \delta H)O^{'\pm}_{l,m}=(\beta^{'\pm}_{l,m})^2 O^{'\pm}_{l,m}$. Since these field amplitudes are dominated by the $O^{\pm}_{l,m}$ amplitudes, these eigenmodes would display the fields of the $LP$ modes \blk(Section 3.1)\blk. Indeed such $LP$ mode-like fields have been observed in numerical simulations using the finite -element solver COMSOL for the slightly elliptical ring-core fibers in \cite{rochelle}. This qualitative agreement gives further credence to our analytic results based on perturbation theory, and suggests additionally that our results, Eqs. 20 and 35, are of a general nature applicable to other fibers as well. Furthermore, the rapid rise in the value of the $2\pi$ walk-off length, predicted by our analytic expressions,  as the topological charge $l$ of the incoming mode is increased (keeping ellipticity $\epsilon$ constant) or the ellipticity $\epsilon$ is decreased (keeping $l$ constant), is also seen in the finite-element calculations in \cite{yue, wang2}. 
\\\\
3) Because the output intensity pattern of the final state  is dominated by the presence of the $OAM_{l,m}$ and $OAM_{-l,m}$ modes in the output mixture, the output intensity pattern is essentially given by $ I(r,\theta)=|(\cos(\pi L/L^{(2\pi)}_{l,m})O_{l,m}+ i \sin(\pi L/L^{(2\pi)}_{l,m}) O_{-l,m})|^2$ (see Eq. 35), which, upon insertion of $O_{l,m} (r,\theta) =(1/\sqrt{N}_{l,m})F_{l,m}(r) e^{il\theta}$ and $O_{-l,m} (r,\theta) =(1/\sqrt{N}_{l,m})F_{l,m}(r) e^{-il\theta}$, reduces to $(1+\sin(2\pi L/L^{(2\pi)}_{l,m})\sin(2l\theta))F^2_{l,m}(r)/N_{l,m}$. This result represents \emph{tilted} \blk$2l$-lobed \blk $LP$ mode patterns (\cite{bhandari, bhandari3, bhandari7}). For example, at $L=L^{(2\pi)}_{l,m}/4$, intensity $ I \propto \cos^2\blk (l\blk(\theta-\pi/4l)\blk)$, \blk implying tilt angles of $\pi/(4l)$. For $l=1$, the tilt angle is $\pi/4$, a result also predicted in \cite{alex}.
%
%
\section{Summary and Discussion}
Working in the weakly guiding approximation, in which the vector modes reduce to products of the spatial OAM modes and the associated polarizations, we have focused on the spatial modes and used the scalar wave equation to develop perturbation theory for  spatial OAM mode mixing and the ensuing OAM modal crosstalk due to  slight ellipticity in a fiber. The developed  perturbation technique  provides insight into the mechanism for mode mixing. A fundamental  mixing selection rule ($\Delta l=\pm2$) makes the problem tractable and  leads to the derivation of analytic expressions for mode-mixing and consequently the crosstalk. The  expressions  include  an explicit formula for the $2\pi$ walk-off length, which is essential in determining the degenerate OAM mode pair mixing. The derived expressions embody  complete mathematical dependence on topological charge, ellipticity, and the length of the traversed fiber, enabling detailed quantitative analyses of the ellipticity-induced crosstalk, and thereby facilitating the study of  optimal fiber parameters for minimal crosstalk. 
\\\\
Scalar perturbation theory is a well-established theory to study perturbations in a given system \cite{landau}. This theory, as developed and presented here, provides a convenient and useful framework to explore the impact of ellipticity. A generic  but vital feature  in the use of this theory is  the negligibility of the  polarization effects, especially the spin-orbit interaction that gives rise to differences in the propagation constants of the vector modes, $HE_{l+1,m}$ and $EH_{l-1,m}$. \blk In this work, we \blk subsequently explore \blk the effect of  spin-orbit interaction to estimate bounds on the values of ellipticity for which the derived results are most valid (accurate). The validity range  expands with increasing values of normalized frequency, $V$, as in a multimode fiber (larger core radius implies relatively smaller spin-orbit interaction), and dwindles as the topological charge is increased, consistent with the fact that the spin-orbit interaction becomes significant at higher values of $l$ and thus cannot be ignored. \blk As a result, the numerical results based on the developed theory are most accurate for low values of topological charge (as determined by the fiber parameters).  \blk We have illustrated the utility of the analytic expressions  and their validity with application to a step-index few mode fiber  and a multimode fiber. 
%
\\\\
Prior work has focused on the  use of computationally intensive finite-element methods that would normally require high levels of precision to obtain detailed relationships with fiber parameters.  Therefore, the  availability of analytic expressions based on the presented scalar theory, along with their utility constraints and the provided insight,  serve as a valuable tool in the analysis and design of fibers for OAM mode propagation.
The technique, as applied to slightly elliptical fibers, is novel and detailed, and, to  our knowledge,  has not been presented  before. This technique also  serves as a precursor to  techniques incorporating groups of degenerate modes as in graded-index fibers. 

\blk
{\large\bf   Appendix}
%
%
%
%
%
\appendix
\numberwithin{equation}{section}
\section {Derivation of the Mixing of the $OAM_{l,m}$ and $OAM_{-l,m}$ Modes due to Fiber Ellipticity}
The derivation here follows closely that for the fiber bend in \cite{bhandari}. The states $OAM_{l,m}$ and $OAM_{-l,m}$ are degenerate. Therefore, there are an infinite number of combinations of these states that satisfy the unperturbed scalar wave equation (Eq. 1).  However,   there is a unique  pair of these  linear combinations, which are orthonormal, and appropriate to the perturbation under consideration.   We now seek these. Due to the Hermitian nature of $H$ and $\delta H$, these linear combinations are related to the above two degenerate states  via a unitary transformation in the $(l,-l)$ subspace. Consequently,  we write
\begin{equation}
O^+_{l,m}=U_{11}O_{l,m}+U_{12}O_{-l,m},
\end{equation}
\begin{equation}
O^-_{l,m}=U_{21}O_{l,m}+U_{22}O_{-l,m},
\end{equation}
where $U_{i,j}$ are the elements of the $2~ x ~2$ unitary matrix $U$. 
%
%
We now replace $O_{ l,m}$ in Eq. 7 with $O^{+}_{l,m}$, and  $O_{- l,m}$ in Eq. 19 with  $O^{-}_{l,m}$, and rewrite the two perturbation  series   in a standard procedure \cite{landau,mw,soliverez}  as
\begin{equation}
O'^{\pm}_{l,m}=O^{\pm}_{l,m}+\sum_{i=1}\sum_{n \ne \pm l,k}a^{\pm (i)}_{(l,m)(n,k)}O_{n,k}, 
\end{equation}
where index $i$ signifies the perturbation order. The two individual series are labeled by $+$ and $-$ signs,  with $n \ne +l $ applying to the former and $n \ne -l$ applying to the latter. The coefficients, $a^{\pm(i)}_{(l,m)(n,k)}$, are defined in the same way as $a_{(l,m)(n,k)}^{(i)}$ (see Eqs. 14 and 15), with the difference, however,  that the matrix elements, $\delta H_{(n,k)(\pm l,m)}=<O_{n,k}|\delta H| O_{\pm l,m}>$ (see Eq. 13) are replaced with   $\delta H^{\pm }_{(n,k)(l,m)}=<O_{n,k}|\delta H| O^{\pm}_{ l,m}>$. Substituting Eqs. A.1 and A.2, we obtain
\begin{equation}
\delta H^+_{(n,k)(l,m)}=<O_{n,k}|\delta H|O^+_{l,m}>=U_{11}\delta H_{(n,k)(l,m)}\delta_{n, l\pm2}+U_{12}\delta H_{(n,k)(-l,m)}\delta_{n,-l\pm 2}
\end{equation}
and
\begin{equation}
\delta H^-_{(n,k)(l,m)}=<O_{n,k}|\delta H|O^-_{l,m}>=U_{21}\delta H_{(n,k)(l,m)}\delta_{n, l\pm2}+U_{22}\delta H_{(n,k)(-l,m)}\delta_{n,-l\pm 2}.
\end{equation}
  Replacing
 $\delta H_{(n,k)(l,m)}$ with Eqs. A.4 and A.5 in the perturbation series, Eq. A.3, transforms it to 
\begin{equation}
O'^{\pm}_{l,m}=O^{\pm}_{l,m}+\sum_{i=1}\sum_{k}a^{(i)}_{(l,m)(n,k)}O^{\pm}_{n,k}
\end{equation}
where
\begin{equation}
O^+_{n,k}=U_{11}O_{n,k}+U_{12}O_{-n,k},
\end{equation}
\begin{equation}
O^-_{n,k}=U_{21}O_{n,k}+U_{22}O_{-n,k},
\end{equation}
and $n=l\pm2i$ (both these values are included in the sum in Eq. A.6).
 Starting from Eq. A.3, we have thus obtained a new form of the series, expressed entirely in terms of linear combinations of the degenerate states. Furthermore, these linear combinations, regardless of the topological charge, are described uniquely by the elements of a single unitary matrix $U$ (to be determined later). The mixing coefficients of $O^{\pm}_{n,k}$ are the same as those in the original perturbation series for $O'_{\pm l,m}$.
\\\\
Consider now
\begin{equation}
(H+\epsilon\delta H)O^{+'}_{l,m}=(\beta'^{+}_{l,m})^2O^{+'}_{l,m}.
\end{equation}
 Writing $(\beta'^{+}_{l,m})^2=\beta^2_{l,m}+\delta \beta'^2_{l,m}$, where $\delta \beta'^2_{l,m}$ is to be  determined along with matrix $U$, and substituting this expression in Eq. A. 9, we obtain
\begin{equation}
(H-\beta_{l,m}^2)O^{+'}_{l,m}=(\delta\beta'^{2}_{l,m}-\epsilon \delta H)O^{+'}_{l,m}.
\end{equation}
Taking the scalar product on the left with $O_{l',m} (l'=\pm l)$, we have
\begin{equation} 
<O_{l',m}|H-\beta_{l,m}^2|O^{+'}_{l,m}>=<O_{l',m}|\delta\beta'^{2}_{l,m}-\epsilon \delta H|O^{+'}_{l,m}>.
\end{equation}
Now $<O_{l',m}|H-\beta_{l,m}^2|O^{+'}_{l,m}>=<(H-\beta_{l,m}^2)O_{l',m}|O^{+'}_{l,m}>$ due to the Hermiticity of $H-\beta_{l,m}^2$.  But $(H-\beta_{l,m}^2)O_{l',m}=0$. Therefore, the LHS of Eq. A.11 equals zero. 
 It follows then
\begin{equation}
<O_{l',m}|\delta\beta'^{2}_{l,m}-\epsilon \delta H|O^{+'}_{l,m}>=0.
\end{equation}
Inserting Eqs. A. 6,  A.1, and A.7 into Eq. A.12, we obtain
\begin{equation}
\begin{split}
&<O_{l',m}|\epsilon\delta H-\delta \beta'^2_{l,m}|(U_{11}O_{l,m}+U_{12}O_{-l,m})> \\
&+\sum_{i=1}^{\infty}\sum_{k}a^{(i)}_{(l,m)(l\pm 2i,k)}<O_{l',m}|\epsilon \delta H -\delta \beta'^2_{l,m}|U_{11}O_{l\pm 2i,k}+U_{12}O_{-(l \pm 2i),k}>=0.
\end{split}
\end{equation}
Now $l'=\pm l$ leads to two linear equations in $U_{11}$ and $U_{12}$:
\\\\
1) \underline{$l'=l$}
\begin{equation}
\epsilon\sum_{i=1}^{\infty}\sum_{k}a^{(i)}_{(l,m)(l\pm 2i,k)}(U_{11} \delta H_{(l,m)(l\pm 2i,k)} +U_{12} \delta H_{(l,m)(-(l \pm 2i),k)})-U_{11}\delta \beta'^2_{l,m}=0.
\end{equation}
\\\\
2) \underline{$l'=-l$}
\begin{equation}
\epsilon\sum_{i=1}^{\infty}\sum_{k}a^{(i)}_{(-l,m)(l\pm 2i,k)}(U_{11} \delta H_{(-l,m)(l\pm 2i,k)} +U_{12} \delta H_{(-l,m)(-(l\pm 2i),k)})-U_{12}\delta \beta'^2_{l,m}=0.
\end{equation}
The above homogeneous equations have a non-trivial solutions if and only if the determinant of the coefficients of $U_{11}$ and $U_{12}$ is zero. Invoking the real and symmetric nature of  $\delta H$ matrix elements and the fact that the mixing coefficients for $l$ and $-l$ cases are identical in values (see Section 3), it is easy to see that the  corresponding matrix  is of the form, $M-\delta \beta'^2_{l,m}I$, where matrix $M$ is  symmetric (with diagonal elements equal) and $I$ is the $2 ~x~ 2$  identity matrix. For degeneracy to be broken between the $OAM_{l,m}$ and $OAM_{-l,m}$ modes, the off diagonal elements of the $M$ matrix must be non-zero. Examining the coefficient of $U_{12}$ in Eq. A.14, which is an off diagonal element of the $2~x~2 $ M matrix, we can set $ \delta H_{(l,m)(-(l \pm 2i),k)}= \delta H_{(-l,m)(l \pm 2i,k)}=\delta H_{(l\pm 2i,k)(-l,m))}$, using the symmetric properties of the $\delta H$ matrix elements (see Section 3). Invoking the $|\Delta l|=2$ rule, this element is non zero only if $i=l-1$ (this follows from choosing the - sign in the index $l \pm 2i$; the + sign gives an infeasible solution).  The value of this off diagonal term, denoted $\gamma_{l,m}$, is given by
%
\begin{equation}
\gamma_{l,m}=\sum_k\epsilon~ a^{(l-1)}_{(l,m)(-l+2,k)}\delta H_{(-l+2,k)(-l,m)}.
\end{equation}
Expanding out the $a^{(l-1)}_{(l,m)(-l+2,k)}$ coefficient as a product of $l-1~\delta H$ elements and the $l-1$ accompanying propagators (see Section 3), we obtain the form
\begin{equation}
\begin{split}
\gamma_{l,m}=&(\epsilon)^{l}\sum\frac{\delta H_{(l,m)(l-2,n)}\delta H_{(l-2,n)( l-4,p)}....}
{(\beta_{l,m}^2-\beta_{l-2,n}^2)(\beta_{l,m}^2-\beta_{l-4,p}^2)....}\\
&\frac{....\delta H_{(-l+4,q)(-l+2,\blk k\blk)}\delta H_{(-l+2,k)(-l,m)}}{....(\beta_{l,m}^2-\beta_{-l+4,q}^2)(\beta_{l,m}^2-\beta_{-l+2,k}^2)}~ (l>1).
\end{split}	
\end{equation}
Identical result follows if we examine the other off diagonal term of the M matrix (coefficient of $U_{11}$ in Eq. A.15). For $l=1$
$\gamma_{l,m}= \epsilon\delta H_{(l,m)(-l,m)}~(l=1)$.
Furthermore, we observe from Eq. A.14 that in the lowest order, which corresponds to $i=1$, the coefficient  of $U_{11}$, denoted $\kappa_{l,m}$, is given by 
\begin{equation}
\kappa_{l,m} = \sum_k\epsilon a^{(1)}_{(l,m)(l+ 2,k)}\delta H_{(l,m)(l+ 2,k)} + \sum_{k'} a^{(1)}_{(l,m)(l- 2,k')} \delta H_{(l,m)(l- 2,k')}.
\end{equation}
Substituting Eq. 14, we find
\begin{equation}
\begin{split}
\kappa_{l,m}=&\epsilon^2\bigg(\sum_k\frac{\delta H_{(l,m)(l+2,k)}\delta H_{(l+2,k)(l,m)}}{(\beta_{l,m}^2-\beta_{l+2,k}^2)}\\
&+\sum_{k'}\frac{\delta H_{(l,m)(l-2,k')}\delta H_{(l-2,k')(l,m)}}{(\beta_{l,m}^2-\beta_{l-2,k'}^2)}\bigg),
\end{split}
\end{equation}
which is identified with $\beta^{2(2)}_{l,m}$ (see Eq. 17); this is not surprising, since $\kappa_{l,m}$ is the diagonal term of the $2~x~2$ matrix $M$ in the $(l,-l)$ subspace. It is the nonzero nature of the off diagonal term, $\gamma_{l,m}$, which breaks the degeneracy.
Matrix  M can now be written as
\begin{equation}
M=
\begin{bmatrix}
\beta^{2(2)}_{l,m}&\gamma_{l,m}\\
\gamma_{l,m}& \beta^{2(2)}_{l,m}
\end{bmatrix}.
\end{equation}
The eigenvalue equation is  $(M-\delta \beta'^2_{l,m}I)\phi= 0$, where $ \delta\beta'^2_{l,m}$ is the eigenvalue of matrix $M$, and $\phi$ (a  $2~x~1$ column vector) the corresponding eigenvector.  Setting the determinant of ($M-\delta \beta'^2_{l,m}I$) to zero then yields the two eigenvalues:
\begin{equation}
\delta \beta'^{\pm2}_{l,m}=\beta^{2(2)}_{l,m}\pm\gamma_{l,m}.
\end{equation}
 That is,
\begin{equation}
(\beta^{'\pm }_{l,m})^2=\beta^2_{l,m}+\beta^{2(2)}_{l,m}\pm\gamma_{l,m}.
\end{equation}
The corresponding eigenvectors are $\frac{1}{\sqrt{2}}\begin{bmatrix}
1\\
1
\end{bmatrix}$ and  $\frac{1}{\sqrt{2}}\begin{bmatrix}
1\\
-1
\end{bmatrix}$, which are identified with the two column vectors of the $U$ matrix.   That is,  $U_{11}=\frac{1}{\sqrt{2}}$, $U_{12}=\frac{1}{\sqrt{2}}$, $U_{21}=\frac{1}{\sqrt{2}}$, and $U_{22}=-\frac{1}{\sqrt{2}}$. The unitary matrix $U$ is independent of $l$. Inserting   the numerical values of the appropriate $U_{ij}$ elements  in Eqs. A.1 and A.2, we obtain  $O^+_{l,m}=\frac{1}{\sqrt{2}}(O_{l,m}+O_{-l,m})$ and $O^-_{l,m}=\frac{1}{\sqrt{2}}(O_{l,m}-O_{-l,m})$.    Similar insertions in Eqs. A.7 and A.8 yield $O^+_{n,k}=\frac{1}{\sqrt{2}}(O_{n,k}+O_{-n,k})$ and $O^-_{n,k}=\frac{1}{\sqrt{2}}(O_{n,k}-O_{-n,k})$, where $n=l+2i, i>0$. As a result, we now know the entire perturbation series, Eq. A.6.  
\\\\
Below, we indicate a few example forms  for  $\gamma_{l,m}$, using Eq. A.17.
For $l=2$,
\begin{equation}
\gamma_{2,m}=\epsilon^{2}\sum_{i}\frac{(\delta H_{(2,m)(0,i)})(\delta H_{(0,i)(-2,m)})}{\beta^2_{2,m}-\beta^2_{0,i}}.
\end{equation}
For $l=3$ 
\begin{equation}
\gamma_{3,m}=\epsilon^{3}\sum_{i,j}\frac{(\delta H_{(3,m)(1,i)})(\delta H_{(1,i)(-1,j)})(\delta H_{(-1,j)(-3,m)})}{(\beta^2_{3,m}-\beta^2_{1,i})(\beta^2_{3,m}-\beta^2_{1,j})},
\end{equation}
where we have used the fact that $\beta^2_{-1,j}=\beta^2_{1,j}$.
\\\\
We can similarly write down expressions for higher values of $l$. 
\section{Derivation of the Bounds on Ellipticity}
The spin-orbit ($SO$) interaction correction to the scalar propagation constant, denoted  $\delta\beta^{2(SO)}_{(l,m)}$,  is  of the order of $l\Delta/a^2$  \cite{snyder, volyar, bhandari4}. These corrections  are of opposite sign for the spin-orbit aligned case (e.g., $+l, S=+1$) and spin-orbit anti-aligned case (e.g.,$-l, S=+1$), where $S$ denotes the spin of the OAM mode, with $S=\pm1$ corresponding to left/right circular polarization.  \blk{Consequently}\blk, we expect these  differences to manifest as unequal diagonal elements in the $2~x~2$ matrix considered in Section 3.1 and Appendix A.  Therefore, to maintain the validity of  our results, 
\blk
 we must ensure that these differences 
\blk due to the SO effects are much smaller than the  splitting of the degenerate spatial modes, $2\gamma_{l,m}$ due to ellipticity $\epsilon$ (Eq. 22), \blk  i.e., the $(l,-l)$  degeneracy is broken primarily through the ellipticity effects. \blk By examining the analytic form for $\gamma_{l,m}$ ( Eq. 22), we  estimate the condition under which this may hold.\blk 
%
%
\\\\
 From Eq. 13, we see that $\epsilon \delta H$ is of the order of $\epsilon \Delta k^2n_1^2$. 
The magnitude of $\gamma_{l,m}$ in Eq. 22 is determined by $l$ such factors in the numerator and $l-1$ factors involving propagation constant differences: $\beta^2_{l',m'}-\beta^2_{l'\pm2,m''}\approx 2kn_1(\beta_{l',m'}-\beta_{l'\pm2,m''})$. We now recall that the total number of $(l,m)$ pairs in a fiber (not counting polarization states and the degenerate counterparts) is approximately $\alpha V^2/8$ (for large $V$) \cite{snyder, buck}, where $\alpha$ varies from  0.5 for a parabolic fiber to 1 for a step-index fiber. This implies  an average mode spacing of $k(n_1-n_2)/(\alpha V^2/8)$ that can for practical purposes for few mode (small $V$) as well as multimode (large $V$)  fibers be represented approximately by $k(n_1-n_2)/V$; $V$ is the normalized frequency equal to $ka(n_1^2-n_2^2)^{1/2}=kan_1(2\Delta)^{1/2}$.  Thus, 
%
$\gamma_{l,m} \sim (\epsilon\Delta k^2 n_1^2)^l/(2k\blk ^2\blk n_1(n_1-n_2)/V)^{l-1}= \Delta k^2 n_1^2 \epsilon^l (V/2)^{l-1}$.
%
Imposing the condition  $
 \delta\beta^{2(SO)}_{(l,m)}<< \gamma_{l,m}$ now leads to the inequality
\begin{equation}
\Big(\frac{l}{k^2a^2n_1^2(V/2)^{l-1}}\Big) << \epsilon^l.
\end{equation}
%
%
We further derive an upper bound from the  requirement that the magnitude of the first order amplitudes $a^{(1)}_{(l,m)(l',m')}$ (Eq. 14) be much smaller than unity\cite{landau} for results to be valid in first order perturbation. From the above discussion, this implies $\epsilon\Delta k^2n_1^2<<2k^2n_1(n_1-n_2)/V$, yielding
\begin{equation}
 \epsilon << 2/V.
\end{equation}
 Combining Eqs. B.1 and B.2, we obtain
\begin{equation}
\Big(\frac{lV}{2k^2a^2 n_1^2}\Big)^{1/l}\Big(\frac{2}{V}\Big) << \epsilon << \frac{2}{V}.
\end{equation}
\blk We note here that the left hand side inequality becomes less stringent (compared to Eq. B.1) for higher values of $l$.

\end{document}